\documentclass[ss, noshowframe]{imsart}

\RequirePackage{amsthm,amsmath,amsfonts,amssymb}
\RequirePackage[authoryear]{natbib}
\RequirePackage[colorlinks,citecolor=blue,urlcolor=blue]{hyperref}
\RequirePackage{graphicx}

\startlocaldefs
\theoremstyle{plain}

\newtheorem{theorem}{Theorem}[section]
\newtheorem{lemma}[theorem]{Lemma}
\newtheorem{proposition}{Proposition}
\theoremstyle{definition}

\theoremstyle{remark}

\newcommand{\bX}{\mathbf{X}(s)}

\newcommand{\bY}{\mathbf{Y}(s)}

\newcommand{\bZ}{\mathbf{Z}(s)}

\newcommand{\eye}{\boldsymbol{I}}

\endlocaldefs

\begin{document}
\begin{frontmatter}
\title{Spatial Confounding: A review of concepts, challenges, and current approaches}
\runtitle{Spatial Confounding: A review}

\begin{aug}
\author[A]{\fnms{Isaque}~\snm{Vieira Machado Pim}\ead[label=e1]{isaque.pim@fgv.br}\orcid{0009-0000-0200-8530}},
\author[B]{\fnms{Marcos}~\snm{Oliveira Prates}\ead[label=e3]{marcosop@gmail.com}\orcid{0000-0001-8077-4898}}
\and
\author[A]{\fnms{Luiz}~\snm{Max Carvalho}\ead[label=e2]{lmax.fgv@gmail.com}\orcid{0000-0001-5736-5578}}
\address[A]{School of Applied Mathematics, Getulio Vargas Foundation, Brazil\printead[presep={,\ }]{e1,e2}}

\address[B]{Department of Statistics, Universidade Federal de Minas Gerais, Brazil\printead[presep={,\ }]{e3}}
\runauthor{Pim et al.}
\end{aug}

\begin{abstract}
Spatial confounding is a persistent challenge in spatial statistics, influencing the validity of statistical inference in models that analyze spatially-structured data.
The concept has been interpreted in various ways but is broadly defined as bias in estimates arising from unmeasured spatial variation.
In this paper we review definitions, classical spatial models, and recent methodological advances, including approaches from spatial statistics and causal inference.
We provide an unified view of the many available approaches for areal as well as geostatistical data and discuss their relative merits both theoretically and empirically with a head-to-head comparison on real datasets.  
Finally, we leverage the results of the empirical comparisons to discuss directions for future research.
\end{abstract}



\end{frontmatter}
\section{Introduction}

Spatially referenced data emerge in many applied areas, including environmental science, ecology and  epidemiology.
The latter constitutes a rich set of applications, such as understanding the spread of infectious disease, pointing to determinants of increased cancer rates, and investigating the association between exposure to fine particles and childhood development \citep{2000_Elliott, 2006_Reich_Biometrics, 2019_Papadogeorgou_Biostatistics}.
When used to draw conclusions for spatially referenced data, standard regression models can result in spatial dependence in the residuals and invalidate the independence assumption \citep{1993_Cressie, 2015_Banerjee_book}.
A common remedy is to include a flexible spatial effect (e.g., Gaussian process, splines, CAR/SAR) so that remaining dependence is absorbed by a spatial effect and uncertainty is better calibrated, improving model inference
\citep{2004_Waller}.

However, when covariates themselves vary smoothly over space, the spatial random effect can ``compete'' with those covariates for the same signal.
The resulting collinearity makes it difficult to distinguish covariate effects from the latent spatial field; coefficient estimates may be attenuated, their standard errors inflated, and sometimes their sign or significance can change relative to a non-spatial model.
This lack of identifiability between fixed effects and the spatial random effect is known as \emph{spatial confounding} \citep{2006_Reich_Biometrics}.
Since its recognition, an active and evolving literature has clarified when spatial confounding arises and is continually developing methods to mitigate it.

Methods to alleviate spatial confounding have already been applied to a wide array of areas: joint species modeling \citep{joint, joint2, joint3}; joint modeling of cancer \citep{azevedo2021mspock}; incidence of poverty \citep{poverty}; prevalence of psychosis \citep{psycho}; survival data \citep{azevedo2023alleviating}; Student's abilities and school facilities \citep{school,flores2021spatial}; hurdle models \citep{2020_Pereira_JRSSA}; and group therapy studies \citep{2016_Paddock_StatMed}. 

Despite a rapidly expanding toolbox for mitigating spatial confounding, the evidence is fragmented -- definitions, estimands, tuning choices, and performance metrics vary across studies -- so results do not cumulate and practitioners lack clear guidance about which method works best under which conditions.
In this paper, we provide a thorough review of methods dedicated to alleviating spatial confounding: for each, we summarize the formulation, the original contributions, reference applications to guide practitioners, and current implementations.
We then place leading approaches on equal footing and conduct a large, standardized comparison study across diverse real datasets, harmonizing targets of inference and evaluation metrics to enable a fair, side-by-side comparison.
The outcomes are practical scenario-based recommendations that clarify the bias–variance trade-offs of competing methods and help readers choose an appropriate strategy for their data.

The remainder of paper is organized as follows: In Section \ref{sec:background}, we present the background and notation for spatial confounding.
Section \ref{sec:filter} reviews early works on spatial statistics,exploring spatial filtering methods.
Restricted spatial regression (RSR) models are discussed in Section \ref{sec:RSR} for both areal and geostatistical data.
We also discuss Transformed Gaussian Markov random fields (TMGRF) and the consequences of orthogonal smoothing.
In Section \ref{sec:scale_confusion}, the scale of confounding is introduced, along with adjustment methods such as spatial basis adjustment, geoadditive structural equation models (gSEM), the Spatial+ approach, spectral adjustment, correlating Gaussian random fields, and regularized spline functions.
An analytical framework for confounding is developed in Section \ref{sec:analytical_frame} and Section \ref{sec:causal} presents a perspective from causal inference, covering topics such as propensity score adjustment, distance-adjusted propensity score matching and double machine learning for spatial confounding.
Section \ref{sec:case_studies} contains, to the best of our knowledge, the broadest empirical study of spatial confounding methods in the literature.
Finally, Section \ref{sec:conclusion} presents conclusions and future directions.

\section{Background and notation}
\label{sec:background}


Consider a spatial domain $\mathcal{D}$ where we observe data at locations $s \in \mathcal{D}$.
This spatial domain can take different forms depending on the nature of the data; for instance, in a geostatistical setting $\mathcal{D} \subset \mathbb{R}^2$, where each location $s \in \mathcal{D}$ represents a coordinate pair in a continuous space.
For areal data on the other hand, the spatial domain consists of a finite set of discrete regions, $\mathcal{D} = \{1, \dots, n\}$, where each $s$ corresponds to a predefined spatial unit such as a county, city, or country.
In this framework, spatial relationships are often described using adjacency structures or neighborhood matrices to capture the spatial dependencies between regions \citep{2015_Banerjee_book}.
We use $\boldsymbol{W}$ to refer to a spatial weight matrix throughout the text.

Spatial regression models are commonly used when residual spatial dependence persists after accounting for observed variables \citep{1993_Cressie}.
This residual dependence can be attributed to an unobserved spatially structured variable \citep{2004_Waller}.  
This in turn motivates the following data-generating process, which underpins various methods for addressing confounding:  
\begin{equation}  \label{Spatial_Generating_Model}  
    \mathbf{Y}(s) = \beta_0 + \beta_X \mathbf{X}(s) + \mathbf{Z}(s) + \mathbf{\varepsilon}(s).
\end{equation}  
Here, \(\mathbf{Y}(s)\) represents the outcome variable, \(\mathbf{X}(s)\) denotes the exposure of interest, and \(\mathbf{Z}(s)\) captures an unobserved (unmeasured) spatially structured variable. The term \(\mathbf{\varepsilon}\) accounts for random error.
The primary goal is to accurately estimate the main effect \(\beta_X\).
However, since \(\mathbf{Z}(s)\) is unobserved, its omission can introduce bias in the estimation of \(\beta_X\).
This issue is known as confounding, and when the unobserved variable \( \mathbf{Z}(s) \) exhibits spatial structure, it is specifically referred to as \textbf{spatial confounding}.  
Methods designed to address spatial confounding aim to mitigate the estimation bias in \(\beta_X\), ensuring more reliable inference. Following the approach of \citet{2024_Khan_preprint}, we differentiate between the data-generating model \eqref{Spatial_Generating_Model} and the analysis models used for inference.
These models are, broadly:  
\begin{align}  
    \textbf{Non-Spatial Models}: &\quad \mathbf{Y}(s) = \beta_0 + \boldsymbol{\beta}_X^{NS}\,\mathbf{X}(s) + \boldsymbol{\varepsilon}(s), \label{eq:NSmodel}\\  
    \textbf{Spatial Models}: &\quad \mathbf{Y}(s) = \beta_0 + \boldsymbol{\beta}_X^{S}\,\mathbf{X}(s) + \boldsymbol{\gamma}(s) + \boldsymbol{\varepsilon}(s), \label{eq:Smodel}\\  
    \textbf{Spatially Adjusted Models}: &\quad \tilde{\mathbf{Y}}(s) = \beta_0 + \boldsymbol{\beta}_X^{AS}\,\tilde{\mathbf{X}}(s) + \boldsymbol{\gamma}(s) + \boldsymbol{\varepsilon}(s). \label{eq:ASmodel}
\end{align}  
 
Here, \(\varepsilon(s)\) represents independent and identically distributed (i.i.d.) noise with variance \(\sigma^2\), while \(\gamma(s)\) represents a term commonly included in spatial models to account for spatial dependence.
For multivariate covariates, the slope parameter is vector-valued, so \(\beta_X\) is replaced by \(\boldsymbol{\beta}\in\mathbb{R}^p\) and the linear term is written as \(\mathbf{X}(s)^\top\boldsymbol{\beta}\).
In areal data, \(\gamma(s)\) typically represents a spatially structured random effect, often modeled using spatial linear mixed models such as Conditional Autoregressive \citep[CAR,][]{1974_Besag_JRSSB} or Simultaneous Autoregressive \citep[SAR,][]{1998_Anselin} models and many others \citep[e.g.,][]{1991_Besag_AISM,prates2012dengue,cruz2023inducing}.
These models impose spatial correlation based on the neighborhood structure, ensuring that geographically closer areas exhibit stronger dependencies.  
In geostatistical data, \(\gamma(s)\) generally represents a smooth spatial surface, making the model a partial linear model.
Alternatively, it can be interpreted as a realization of a continuous Gaussian random field, commonly specified using a Matérn covariance function.
The Matérn field provides a flexible framework for modeling spatial correlation, allowing control over smoothness, variance, and range of spatial dependence \citep{matern2013spatial}.
In the adjusted model, the transformed variables \(\tilde{\mathbf{Y}}(s)\) and \(\tilde{\mathbf{X}}(s)\) differ from the original \(\mathbf{Y}(s)\) and \(\mathbf{X}(s)\) due to the spatial adjustment, which alters the spatial structure of the input data used in the models.

We now review the currently available literature on spatial confounding, highlighting the advances and seeking to fill in the gaps between papers in order to provide a coherent view of the challenges and promising directions for the area.

\section{Early work on Spatial Statistics}
\label{sec:filter}

In order to contextualize the main methodological approaches to spatial confounding, it pays to consider the history of Spatial Statistics as a whole.
In a recent discussion, \citet{2024_Donegan_GA} highlights the parallels between the growing literature on spatial confounding and the well-established body of work on spatial autocorrelation.

Despite these historical insights, many classical models from the spatial autocorrelation literature are often overlooked when evaluating methods for addressing spatial confounding.
In particular, key contributions from the econometrics literature are frequently absent from these discussions.
For instance, \citet{2009_LeSagePace} dedicates a section to the study of omitted variable bias, illustrating how spatial regression models can mitigate this bias more effectively than ordinary least squares (OLS) methods.

A key method in the spatial statistics literature that directly relates to spatial confounding is the idea of spatial filtering (SF).
SF aims at improving the robustness and accuracy of spatial data analysis by decomposing a spatial variable into three distinct components: a deterministic trend, a spatially structured random component, and random noise \citep{2014_Griffith}.
This decomposition helps isolate the effects of spatial dependence, allowing for more reliable statistical inference.  

Formally, the goal is to decompose a geographic variable \( \mathbf{Y} \) as $\mathbf{Y} = \mathbf{Y}^*(s) + \nu(s)$,
where \( \nu(s) \) represents the spatially autocorrelated component, and \( \mathbf{Y}^*(s) \) is the independent (or spatially-filtered) component.
The key idea behind this decomposition is to remove spatial dependence from \( \mathbf{Y} \), ensuring that standard statistical methods can be applied to \( \mathbf{Y}^*(s) \) without violating assumptions of independence.  

Spatial filtering techniques include eigenvector-based methods (e.g., eigenvector spatial filtering), which construct spatial basis functions from the eigendecomposition of a connectivity matrix \citep{Griffith_2003, 2014_Griffith, 2007_TiefelsdorfGriffith_EigenvectorFiltering}.
Although Fourier-domain filtering is less commonly used as an explicit adjustment device in spatial regression, frequency-domain methods are standard in digital image processing and remote sensing, where manipulating the spatial-frequency spectrum is routinely used to suppress high-frequency (small-scale) noise and remove periodic artifacts in imagery \citep{RichardsJia1999Fourier, Schowengerdt2007RemoteSensing, GonzalezWoods2022DIP, whittle1954stationary}.
Once the spatial dependence is accounted for, inference can be performed on \( \mathbf{Y}^* \) using traditional statistical models, such as OLS regression, without concerns of spatial autocorrelation distorting results.

The Spatial Lag Model \citep[SLM;][]{1998_Anselin} serves as a foundational example of spatial filtering.
By modeling the outcome as $\mathbf{Y} = \rho \boldsymbol{W} \mathbf{Y} + \mathbf{X} \boldsymbol{\beta} + \boldsymbol{\varepsilon}$, the transformation $(I - \rho \boldsymbol{W})\mathbf{Y} = \mathbf{X} \boldsymbol{\beta} + \boldsymbol{\varepsilon}$ effectively applies a high-pass filter to the outcome variable.
This operation subtracts local means—represented by the spatially weighted average $\boldsymbol{W}\mathbf{Y}$—to remove spatial trends and isolate the non-spatial component of $\mathbf{Y}$ for inference on $\boldsymbol{\beta}$ \citep{2008_Pace_SSRN}.
Failing to account for this dependence can lead to severe omitted variable bias in OLS estimates, particularly when spatial autocorrelation exists across regressors and disturbances \citep{2008_Pace_SSRN}.
This bias often leads to incorrect empirical conclusions; for instance, SLM corrections have been shown to reverse counter-intuitive OLS results in environmental hedonic pricing \citep{2005_Brasington_RSUE} and retail sales modeling \citep{2005_Lee_JREFE}, yielding estimates that align more closely with economic theory.

\subsection{Spatial filtering methods}
\label{sec:sf_methods}
Moving on to key spatial filtering methods, two 
approaches are presented in \citet{2002_GetisGriffith_GA}. 
The first method 
involves a direct transformation of the variable using its Local Indicator of Spatial Association \citep[LISA,][]{anselin1995local}, specifically the Getis-Ord \( G_i(d) \) statistic \citep{getis1991spatial}.
The Getis-Ord statistic measures the local spatial autocorrelation of a variable at a given location \( i \) by assessing the concentration of high or low values within a specified distance \( d \). 
Formally, \( G_i(d) \) is given by: $G_i(d) = \frac{\sum_{j} w_{ij}(d) x_j}{\sum_{j} x_j}$,
where  \( x_j  = \mathbf{X}(s_j)\) is the value of the variable at location \( s_j \), \( w_{ij}(d) \) is a spatial weight that defines the neighboring structure within distance \( d \).
In the case of a binary adjacency matrix $W$ for areal data, $w_{ij} = (W)_{ij}$ that is 1 if the areas are neighbors and 0 otherwise.
The denominator ensures normalization based on the sum of all observations.
This expression is the \textit{observed} value of the statistic for location \( i \).
The \textit{expected} value of the statistic for location \( i \) is of the form \( W_i / (n-1) \), where $W_i$ is the sum of binary weights in row $i$.
The filtering transformation of the observed data values \( \mathbf{X}(s_i) = x_i\) applied is then: $
x_i^* = x_i \frac{\text{Expected}_i}{\text{Observed}_i}, \quad \text{i.e.} \quad x_i^* = x_i \left( \frac{W_i}{n-1} \right) / G_i(d).$

The second spatial filtering method , introduced by Griffith, is known as eigenvector spatial filtering (ESF) or Moran eigenvector filtering.
This approach leverages the computational formula for Moran's \( I \) statistic \citep{1950_Moran_Biometrika} to decompose spatial autocorrelation into orthogonal and uncorrelated spatial components. 
Moran’s \( I \) measures global spatial autocorrelation, capturing the degree to which nearby values of a variable are similar.
The ESF method decomposes Moran’s \( I \) statistic into spatial eigenvectors, which represent uncorrelated spatial patterns.
These eigenvectors serve as latent variables that capture systematic spatial dependence in the dataset. 

The eigenvectors used for spatial filtering are derived from a modified spatial weights matrix $\boldsymbol{W}^*$, which appears in the numerator of the Moran coefficient \citep{2019_DanielChunLi}:
\begin{equation}
\label{eq:modified_SWM}
\boldsymbol{W}^* = \boldsymbol{M} \boldsymbol{W} \boldsymbol{M}, \; \text{where} \; \boldsymbol{M} = \mathbf{I} - \frac{\mathbf{1} \mathbf{1}^T}{n}.
\end{equation}

Here \( \boldsymbol{W} \) is the spatial weights matrix defining the spatial relationship among observations, \( \mathbf{I} \) is the identity matrix,  \( \mathbf{1} \) is a vector of ones,  \( \boldsymbol{M} \) is a centering matrix that ensures the eigenvectors capture spatial variations while removing non-spatial trends.  

Once the eigenvectors are extracted from the transformed matrix \( \boldsymbol{M} \boldsymbol{W} \boldsymbol{M} \), they represent a set of independent spatial patterns present in the data.
The next step is to select a subset of eigenvectors based on their Moran’s \( I \) values.
Typically, eigenvectors exceeding a certain Moran’s \( I \) threshold are chosen to account for significant spatial dependence \citep{2013_Hughes_JRSSB}. 
Once a subset of eigenvectors are selected, they are regressed against the dependent variable \( \mathbf{Y} \)  in a stepwise fashion under a multiple regression model: $
\mathbf{Y} = \sum_{k=1}^{m} \alpha_k E_k + \varepsilon$, where \( E_k \) are the selected spatial eigenvectors,  and \( \alpha_k \) are their corresponding regression coefficients.
The residuals of this regression correspond to the spatially filtered version of \( \mathbf{Y} \), which are largely free of spatial autocorrelation, $ \mathbf{Y}^*(s) = \mathbf{Y} - \sum_{k=1}^{m} \alpha_k \mathbf{e}_k$.

Thus, the filtered variable \( \mathbf{Y}^*(s) \) can be used in traditional statistical models without violating the assumption of independence.  
For more details, see \cite{Griffith_2003} and \cite{2019_DanielChunLi}.
A Bayesian formulation for the spatial filter was proposed by \citet{hughes2017spatialregressionbayesianfilter} and by \citet{2020_Donegan_SS}.
The structure is the same as in the Moran eigenvector filtering, where \citet{hughes2017spatialregressionbayesianfilter} formulates Bayesian eigenvector filtering by augmenting the regression with a selected subset of Moran eigenvectors and assigning spherical priors to the corresponding coefficients, whereas \citet{2020_Donegan_SS} retains the full eigenvector basis but places regularizing shrinkage priors on the coefficients to control complexity.
The key advantage of Donegan’s approach is that it replaces \textit{ad hoc} eigenvector selection with principled regularization and model averaging, propagating uncertainty about the spatial filter into inference for the coefficients.

\section{Restricted Spatial Regressions}
\label{sec:RSR}

Significant attention has been directed toward the bias that arises when incorporating spatial smoothing into regression frameworks in the disease mapping literature.
A seminal example is provided by \citet{1993_Clayton_IJE}, who analyzed lung cancer incidence in Sardinia and described this phenomenon as "confounding by location".
Building on these observations, \citet{2006_Reich_Biometrics} reasoned on this issue, framing it as a form of collinearity where the spatial random effects compete with the covariate matrix $\mathbf{X}$ to explain the same variation.
To mitigate this, they proposed Restricted Spatial Regression (RSR), which constrains the spatial random effects to the orthogonal complement of the column space of $\mathbf{X}$.
This restriction ensures the spatial component captures only residual geographic structure, preventing it from attenuating or destabilizing inference on the regression coefficients.
The work by \citet{2006_Reich_Biometrics} proved foundational, motivating a recent spur in the spatial confounding literature.
In the following section, we detail the core models of the RSR literature, highlighting their respective strengths and inferential trade-offs.
We first examine the foundational RSR framework for areal data, followed by its extension to continuous spatial processes.
Finally, we discuss recent critiques of RSR, specifically addressing how the orthogonal constraint may lead to under coverage of uncertainty intervals.

\subsection{Restricted Spatial Regressions on areal data}
\label{sec:rsr_area}
\citet{2006_Reich_Biometrics} take a Bayesian approach to the study of the impact of an ICAR prior on the estimation of $\beta_X$. The authors employ the ICAR model as a spatial component in the following formulation:
\begin{eqnarray} \nonumber
\mathbf{Y} | \boldsymbol{\beta}, \boldsymbol{\gamma}, \tau_{\epsilon} \sim \mathcal{N}(\mathbf{X} \boldsymbol{\beta} + \boldsymbol{\gamma}, \tau_{\epsilon} \mathbf{I}), \; \text{\textbf{with}} \; 
\boldsymbol{\gamma} | \tau_{\gamma} \sim \mathcal{N}(\mathbf{0}, \tau_{\gamma} \mathbf{Q}),
\end{eqnarray}
where $\boldsymbol{\gamma}$ is the ICAR spatial effect, $\mathbf{Q}$ is the precision matrix for the ICAR, $\tau_{\epsilon}$ and $\tau_{\gamma}$ are precision hyperparameters related to the Gaussian observations and the ICAR, respectively.
In the case of spatial linear regression, it is possible to analytically calculate the marginal mean integrating out the latent effect as:
\begin{eqnarray} \nonumber
\mathbb{E}(\boldsymbol{\beta} | \tau_{\epsilon}, \tau_{\gamma}, \mathbf{Y}) &=& (\mathbf{X}^T \mathbf{X})^{-1} \mathbf{X}^T (\mathbf{Y} - \hat{\boldsymbol{\gamma}}) = \boldsymbol{\beta}_X^{NS} - (\mathbf{X}^T \mathbf{X})^{-1} \mathbf{X}^T \hat{\boldsymbol{\gamma}}.
\end{eqnarray} 
From the result above it is possible to see that spatial models differ in estimation from traditional linear models by a quantity involving the latent effect.
If one could make the latent effect orthogonal to the design matrix, the bias term involving the latent effect would vanish.
To achieve this,  \cite{2006_Reich_Biometrics} propose the following model specification, henceforth called the RHZ model.
First, define the projection matrix onto the column space of $\mathbf{X}$, $\mathbf{P}_X = \mathbf{X}^T(\mathbf{X}^T\mathbf{X})^{-1}\mathbf{X}$, and $\mathbf{P}_X^{\perp} = \eye - \mathbf{P}_X $ its orthogonal complement.
The structured random effect can be decomposed into a component on the span of $\mathbf{X}$, and a component orthogonal to the span of $\mathbf{X}$ by taking 
$\boldsymbol{\gamma} = \boldsymbol{\gamma}^X + \boldsymbol{\gamma}^{\perp} = \mathbf{K} \boldsymbol{\gamma}_1 + \mathbf{L} \boldsymbol{\gamma}_2$.
The proposed solution is to set $\boldsymbol{\gamma}_1$ to zero, neutralizing the components of the random effect in the span of $\mathbf{X}$.
The RHZ model is then formulated as 
\begin{equation*}
    \mathbf{Y}= \mathbf{X}\beta_X + \mathbf{L}\boldsymbol{\gamma}_2 + \varepsilon, \quad  p(\boldsymbol{\gamma}_2|\tau_s) \propto \tau_s^{\kappa} \exp \left( \frac{-\tau_s}{2} \boldsymbol{\gamma}_2^T \mathbf{L}^T\mathbf{QL} \boldsymbol{\gamma}_2 \right).
\end{equation*}

This approach offers a more computationally efficient solution by utilizing \( \mathbf{L}^T \), a \( (n - p) \times n \) matrix, rather than \( \mathbf{P}^{\perp} \), which is an \( n \times n \) matrix.
However, since the number of covariates \( p \) is typically much smaller than the number of areas \( n \), this reduction does not yield a substantial computational advantage.

A better computational relief came with the work of \citet{2013_Hughes_JRSSB}, who noticed that the RHZ model is computationally inefficient and that the model accounts for both positive and negative spatial autocorrelation, arguing that the latter is not desired in spatial applications. 
One of the reasons for the inefficiency is that the precision matrix produced by the projections used by \citet{2006_Reich_Biometrics} is not sparse. 
In addition, negative spatial correlation arises from the fact that in the construction of $L$, the underlying graph is not accounted for.
To address this, \citet{2013_Hughes_JRSSB} define the Moran operator $\mathbf{P}_X^{\perp} \boldsymbol{W} \mathbf{P}_X^{\perp}$ and replace $\mathbf{L}$ by a matrix $\boldsymbol{M}_q$ composed of eigenvectors of the Moran operator. 
Notice how this resembles the modified spatial weight matrix in Equation~\eqref{eq:modified_SWM}.
They showed that the Moran operator retains the spatial patterns of the data better than the columns of $\mathbf{L}$ and it is only necessary to select the $h \ll n$ higher positive eigenvalues of the spectrum of the Moran operator, called attractive eigenvalues, reducing computational burden.

Another major improvement in the computational performance for RSR models was developed by \cite{2019_Prates_BA}.
Even though \citet{2013_Hughes_JRSSB} attempt to improve computational efficiency by reducing the dimension of the problem, much of the effectiveness of ICAR models is that the random fields are very sparse. 
This allows for fast sparse matrix routines to be used \citep{2005_Rue}.
\citet{2019_Prates_BA} try to alleviate confounding by preserving much of the structure and the routines used to fit ICAR models by projecting the graph defined by the study area onto the orthogonal space of the design matrix $\mathbf{X}$, a method they dubbed SPOCK (SPatial Orthogonal Centroid ``K''orrection.). 
With the projected vertices in hand, they construct a sparse precision matrix $\mathbf{Q}^{\perp}$ for analysis.
The first step is to calculate the new set of centroids $c^*$ by projecting $c$ onto the orthogonal space of $\mathbf{X}$, that is, using the projection matrix $\mathbf{P}_X^c$.
After the new set of centroids is produced, a new adjacency matrix can be produced by finding the $k$-nearest-neighbors of each centroid, picking $k$ for each region as the number of neighbors it originally had. 
By doing this, the sparsity of the original adjacency matrix will be kept, thus maintaining the efficiency of sparse methods.
SPOCK can also be applied in a multivariate context \citep{azevedo2021mspock}.

Another key aspect explored in their work is the Type-S error rate in RSR models for areal data, following the approach of \citet{2015_Hanks_Env}.
They define a Type-S error as occurring when a regression parameter is truly zero (\( \beta_X = 0 \) or \( \beta_X^* = 0 \)), yet its $95\%$ symmetric posterior credible interval does not contain zero (see also \cite{2014_Gelman_PsychSci}). 
Their findings reveal an elevated Type-S error rate in RSR models, further reinforcing criticism of their reliability in spatial analyses.

\citet{2021_Nobre_ISR} examine spatial confounding in multilevel models with clustered observations, showing that bias in fixed effects persists even with independent random intercepts due to within-cluster dependencies.
They extend RSR to hierarchical settings, concluding that while RSR reduces bias for cluster-level effects it increases Type-S error rates and may worsen estimation of unit-level coefficients when predictors are correlated.
Their simulations reveal that the magnitude of the bias depends on the spatial scales of covariates and random effects, in agreement with previous work by \cite{2010_Paciorek_StatSci} and \cite{2017_Page_ScandinaviaJS} and which are reviewed in Section~\ref{sec:scale_confusion}.
\citet{2022_Hui_TAS} studies spatial confounding in the context of Generalized Estimating Equations (GEE) and find that spatial confounding can also arise in the setting of GEE and propose a restricted spatial working correlation matrix to correct for confounding.

\subsection{Restricted Spatial Regressions on Geostatistical data}
\label{sec:rsr_geostatistical}

\citet{2015_Hanks_Env} bring the discussion of spatial confounding to geostatistical data and extend the methods presented by \citet{2006_Reich_Biometrics} and \citet{2013_Hughes_JRSSB} to data with continuous support.
To this end, the authors move from a ICAR model specification to a Matérn specification to model autocorrelation.
That is, the model now follows
\begin{equation*}
    \bY = \bX \beta_X + \boldsymbol{\gamma} + \boldsymbol{\varepsilon}, \quad \boldsymbol{\gamma} \sim \mathcal{N}(\mathbf{0}, \mathbf{\Sigma}),
\end{equation*}
with
\begin{equation}
\label{eq:Matern_spec}
\Sigma_{ij} = \sigma^2 C_{\nu}(d_{ij}; \phi) = \sigma^2 \frac{1}{\Gamma(\nu) 2^{\nu-1}} \left( \sqrt{2\nu} \frac{d_{ij}}{\phi} \right)^{\nu} K_{\nu}\left( \sqrt{2\nu} \frac{d_{ij}}{\phi} \right),
\end{equation}
where $d_{ij}$ is the Euclidean distance between the spatial locations of the $i$-th and $j$-th observations, $\sigma^2$ is the partial sill parameter, $\nu$ is the Matérn smoothness parameter, $\phi$ is a range parameter, and $K_{\nu}(\cdot)$ is the modified Bessel function of the second kind \citep[e.g,][]{1993_Cressie}. To perform RSR in a efficient way, they propose a method to constrain $\gamma$ by ``conditioning by Kriging'' \citep{2005_Rue}. 
The orthogonalization is performed \textit{ad hoc} during the MCMC sampling process, by sampling from the Matérn field $\boldsymbol{\gamma} \sim \mathcal{N}(\boldsymbol{\mu}, \boldsymbol{\Sigma})$ with the constraint $\mathbf{X}^T \boldsymbol{\gamma} = 0$.
This can be accomplished by the transformation: $ \boldsymbol{\gamma}^* \sim \mathcal{N}(\boldsymbol{\mu}, \mathbf{\Sigma}), \; \mbox{and}\; \boldsymbol{\gamma} = \boldsymbol{\gamma}^* - \mathbf{\Sigma X} (\mathbf{X}^T \mathbf{\Sigma X})^{-1} \mathbf{X}^T \boldsymbol{\gamma}$.
\citet{2015_Hanks_Env} also contributes to the study of Type-S errors in RSR.
They find through simulations that Type-S errors increase as spatial range (correlation distance) increases and RSR performs poorly when the true model is a SGLMM.
Another exploration of the same scenarios described by \citet{2015_Hanks_Env} is done by
\citet{2017_Hefley_JABES}. 
They do not use RSR, but rather explore regularized models with confounded data adapting the priors for coefficients to include regularization. 
They show empirically that regularization reduces problems with multicollinearity and improves Markov chain Monte Carlo (MCMC) mixing.

\citet{2019_Chiou_SEnvRRA} argue that RSR methods have been developed primarily within a Bayesian framework.
Their contribution to the literature is to extend RSR to the frequentist setting by formulating estimators based on an adjusted version of the generalized least squares (GLS) estimator.
More recently, \citet{2025_Chiou_EnvEcoStat} connect RSR’s projection idea to low-rank geostatistical modeling via Fixed Rank Kriging (FRK): they use an FRK representation of the latent spatial process to estimate the component aligned with the covariate space and then correct inference on $\beta_X$ for spatial confounding.
In this sense, FRK serves as a practical implementation layer for the same spatial-confounding mechanism that motivates RSR, without relying solely on a hard orthogonality constraint.

\subsection{Transformed Gaussian Markov Random Fields}
\label{sec:tgmrf}

Although not strictly an RSR method, transformed Gaussian Markov Random Fields \citep[TGMRF,][]{prates2015transformed} mitigate confounding by structurally isolating the marginal mean specification from the spatial dependence.
While a standard spatial GLMM induces dependence via an additive latent term in the link function (i.e., $g (\mu_i ) = \mathbf{x}_i \boldsymbol{\beta} + \varepsilon_i$), the TGMRF framework models the random mean vector $\boldsymbol{\mu}$ directly via a Gaussian copula.
A random field is denoted $\boldsymbol{\mu} \sim \operatorname{TGMRF}_n(\mathbf{F}_{\boldsymbol{\beta}}, \mathbf{Q}_\rho)$ if it couples independent marginal distributions $\mathbf{F} = (F_1, \dots, F_n)$ with a latent GMRF dependence structure characterized by precision matrix $\mathbf{Q}_\rho$.
The resulting hierarchical formulation is:$$y_i | \mu_i \sim \pi (y_i | \mu_i), \quad \boldsymbol{\mu} \sim \operatorname{TGMRF}_n(\mathbf{F}_{\boldsymbol{\beta}, \boldsymbol{\nu}}, \mathbf{Q}_{\rho}).$$Here, the regression coefficients $\boldsymbol{\beta}$ solely parameterize the marginals $\mathbf{F}$, while the spatial decay $\rho$ is restricted to the copula.

The authors argue that this orthogonality prevents the dependence structure from interfering with the marginal models, effectively alleviating confounding.
\citet{prates2021transformed} formalized this capability, reporting that TGMRFs offer a compromise between standard SGLMMs and RSR methods.
They reduce variance inflation under strong confounding while retaining better type-I error control and interval coverage than restricted approaches, which can become overconfident when confounding is absent.
A spatio-temporal extension of this framework was proposed by \citet{2022_Prates_JRSSC}.
\subsection{Consequences of orthogonal projection of spatial effects}
\label{sec:consequences_orth}

RSR dominated the spatial confounding literature for more than a decade, but has since been argued to be problematic.
\citet{2020_Khan_JASA} and \citet{2022_Zimmerman_TASt} were the first to point out the inconsistencies of the RSR methods and argue against the use of RSR.
Most of the modern accounts already accept these conclusions and reinforce them \citep{2023_Urdagarin_RMC, 2023_Dupont_preprint, 2024_Khan_preprint, 2024_Donegan_GA}.
Even when assuming a data-generating mechanism that agrees with RSR, coverage of the true parameters is worse than a NS model.
This can be checked either with Bayesian methods by means of simulation and analysis of Type-S error as done by \citet{2020_Khan_JASA} or by a frequentist theoretical analysis of the bias of RSR estimators as done by \citet{2022_Zimmerman_TASt}. 

On the Bayesian side, \citet{2019_Prates_BA}, \citet{2015_Hanks_Env} and \citet{2020_Khan_JASA} study the Type-S error for RSR models.
They all reported higher rates of Type-S error for RSR models in all possible scenarios.
\citet{2020_Khan_JASA} brings an analytical explanation for the elevated error rates, where they claim that RSR methods will only capture the regression coefficient if the non-spatial model does.
Furthermore, RSR methods will always have higher rates of Type-S error than the non-spatial model, even if there is no spatial confounding.
However, \citet{2024_Bradley_spatialdeconfoundingreasonablestatistical} argues that these sub-optimal inferential properties arise from a specific interpretation that incorrectly assumes equivalence between confounded and deconfounded effects, and demonstrates that spatial deconfounding remains a reasonable statistical practice when viewed as a reparameterization that produces inferences equivalent to the standard spatial linear mixed model.

On the frequentist side, \citet{2022_Zimmerman_TASt} claimed that deconfounding a spatial linear model by orthogonalization is ``bad statistical practice and should be avoided''.
They come to this conclusion by studying the properties of the MLE estimators of models \eqref{eq:NSmodel}, \eqref{eq:Smodel}, and \eqref{eq:ASmodel}.
When Equation~\eqref{eq:ASmodel} for RSR models is specialized to a frequentist setting we get $\mathbf{Y} = \bX \beta_X + \mathbf{P}_X^c \boldsymbol{\gamma}(s) + \boldsymbol{\varepsilon}(s)$. Thus, assuming the structured component $\boldsymbol{\gamma}(s)$ has variance $\sigma^2 \mathbf{G}$,  the marginal covariance matrix of $\mathbf{Y}$ is given by 
$$\mathbf{\Sigma}_{RSR} = \sigma^2\left[\left(\eye - \mathbf{P}_X \right)\mathbf{G}(\eye - \mathbf{P}_X) + \eye\right]$$.
Then the empirical GLS estimator for the RSR model is 
$$
    \hat{\boldsymbol{\beta}}_{RSR} = \left( \mathbf{X}^T \hat{\boldsymbol{\Sigma}}_{\text{RSM}}^{-1} \mathbf{X} \right)^{-1} \mathbf{X}^T \hat{\boldsymbol{\Sigma}}_{\text{RSM}}^{-1} \mathbf{Y}.$$
They also consider based on \citet{2013_Hughes_JRSSB}, estimators derived from the Moran operator.
The marginal covariance matrix of $\mathbf{Y}$ is given by $\mathbf{\Sigma}_{Moran} = \sigma^2\left[\boldsymbol{M}_q \boldsymbol{M}_q^T \mathbf{G} \boldsymbol{M}_q \boldsymbol{M}_q^T + \eye\right]$.
Then the empirical GLS estimator for the $\beta$ is $
    \hat{\boldsymbol{\beta}}_{Moran} = \left( \mathbf{X}^T \hat{\mathbf{\Sigma}}_{\text{Moran}}^{-1} \mathbf{X} \right)^{-1} \mathbf{X}^T \hat{\mathbf{\Sigma}}_{\text{Moran}}^{-1} \mathbf{Y}$.

More generally deconfounded estimators can be produced by transforming the structured random effect using any matrix derived from the column space of $\mathbf{P}_X^c$. Based on these estimators, they show that the CIs for RSR models are narrower than the the standard CIs from NS models. 
This is in contradiction with a pragmatic thinking in spatial statistics, as described by \citet{2024_Donegan_GA}, that each observation in a spatial context is not as informative as an independent sample.
So any uncertainty quantification that takes this into account should produce wider intervals.
Fortunately for RSR, \citet{2022_Zimmerman_TASt} find that despite its inferiority in estimating fixed effects, RSR does not significantly degrade spatial prediction.
The best linear unbiased predictor (BLUP) of new spatial observations remains identical under both the original spatial model and the deconfounded model.

RSR methods played a crucial role in reigniting discussions on fundamental challenges in spatial regression but have since proven problematic.
The literature moved on to new models, but key findings are still present in many of the works in the area. 
\section{Scale of Confusion}
\label{sec:scale_confusion}

While RSR methods focus on model reformulation, \citet{2010_Paciorek_StatSci} introduced the "scale of confounding" to explain how spatial structure dictates bias from a data-generation perspective.
Under the framework of Equation \eqref{Spatial_Generating_Model}, consider an exposure $\mathbf{X}$ and unmeasured confounder $\mathbf{Z}$ modeled as Gaussian processes: $\mathbf{X} \sim \mathcal{N}(\mathbf{0}, \sigma_x^2 \mathbf{R}(\theta_x))$ and $\mathbf{Z} \sim \mathcal{N}(\mathbf{0}, \sigma_z^2 \mathbf{R}(\theta_z))$, with $\text{Cov}(\mathbf{X}, \mathbf{Z}) = \rho \sigma_x \sigma_z \mathbf{R}(\theta_c)$.
\citet{2010_Paciorek_StatSci} demonstrated that standard spatial regression (kriging or GLS) does not inherently eliminate bias. Specifically, the induced bias in the GLS estimator, $\mathbb{E}(\hat{\beta}_x - \beta_x| \mathbf{X}) = \rho \frac{\sigma_z}{\sigma_x} \beta_z$, is identical to that of OLS.
However, by adopting a multiscale decomposition $\mathbf{X} = \mathbf{X}_c + \mathbf{X}_u$, where $\mathbf{X}_c$ shares a spatial range with $\mathbf{Z}$, the bias becomes scale-dependent:
$\mathbb{E}(\hat{\beta}_X | \mathbf{X}) = \beta_x + c(\mathbf{X}) \rho \frac{\sigma_z}{\sigma_c} \beta_z.$
Here, the function $c(\mathbf{X})$ incorporates the interplay between spatial scales.
This reveals that spatial modeling can actually exacerbate bias if the confounded spatial range is shorter than the unconfounded range; mitigation only occurs when the confounded range is the broader of the two.
This foundational insight underscores that the effectiveness of spatial adjustment depends entirely on the relative scales of the exposure and the unmeasured confounder.

\citet{2017_Page_ScandinaviaJS} further studied the importance of scale for estimation of the parameters but also addresses prediction.
\citet{2024_Narcisi_EnvEcoStat} provides a analysis of confounding through the lens of quadratic forms.

This concept has already been explored in confounding adjustment methods for time series \citep{2004_Dominici_JASA, 2014_Szpiro_Biometrics}.
When the exposure of interest varies at fine spatial scales, smoothing the series can help isolate this variation for inference.
This approach also aligns with the filtering methods discussed in Section \ref{sec:filter}, which remove large-scale spatial variation to facilitate inference using the smoothed variables.


Following the developments of \citet{2010_Paciorek_StatSci} another line of methodology was developed, focusing on adjusting for spatial information in the exposure, or taking into account a joint mechanism behind the generation of the exposure.

\subsection{Spatial Basis Adjustment and Regularization}
\label{sec:spatial_basis}

To address spatial confounding in cohort settings, \citet{2020_Keller_JRSSA} propose decomposing the exposure, covariates, and error term into smooth spatial surfaces via hierarchical basis functions $\{h_1, h_2, \dots\}$ (e.g., splines, Fourier, or wavelets) ordered by increasing resolution.
Let $\mathbf{H}_m$ denote the first $m$ basis functions.
They introduce two adjustment strategies: (i) directly including $\mathbf{H}_m$ as covariates in a semiparametric regression, or (ii) a spatial filtering approach where the exposure $\mathbf{X}$ is projected onto the space orthogonal to $\mathbf{H}_m$.
By utilizing bases ordered by resolution, these methods isolate the fine-scale variation of the exposure for inference while removing the large-scale spatial components typically associated with unmeasured confounders.
A key contribution of \citet{2020_Keller_JRSSA} is providing a formal mechanism for identifying the appropriate spatial scale for confounding adjustment.
Because the bias and variance of the exposure effect are highly sensitive to the number of basis functions used, the authors developed criteria to estimate the "optimal" scale $m$ for different basis types.
By evaluating the trade-off between bias reduction (achieved by including more bases) and loss of efficiency (caused by removing too much exposure variation), their framework allows researchers to objectively determine the resolution at which the exposure provides the most reliable signal for inference.

While \citet{2020_Keller_JRSSA} approach often requires pre-selecting this number of bases $m$ via cross-validation or information criteria, \citet{zaccardi2024regularizedprincipalsplinefunctions} introduce a Bayesian semiparametric model that automates scale selection through regularization.
They approximate the unobserved spatial factor using principal spline basis functions $\mathbf{B}$ and employ spike-and-slab priors to identify the most critical bases: $\beta^s_{j} \mid \gamma_j \sim \gamma_j \mathcal{N}(0, \psi_j^2) + (1 - \gamma_j) \mathcal{N}(0, c_0 \psi_j^2)$. Here, the indicator $\gamma_j \sim \text{Bernoulli}(w)$ determines basis inclusion, while a small $c_0$ encourages sparsity.
This regularized framework—which can be implemented with various prior variances such as the product moment (pMOM) prior \citep{2012_Johnson_JASA}—provides a data-driven mechanism to select the appropriate spatial resolution for confounding adjustment without manual tuning of the basis rank.

\subsection{Spatial+}

Spatial+ \citep{2022_Dupont_Biometrics} is a two-stage approach to spatial confounding that targets the spatial structure of the exposure rather than altering the spatial effect in the outcome model.
It first decomposes the covariate into spatial and non-spatial components and then performs inference using the non-spatial remainder, while retaining a flexible spatial term in the response model.

One again, let $\mathbf{X}(s)$ being the covariate and $f(\mathbf{s})$ a smooth function of space that captures the spatial effect. 
Spatial+ then decomposes $X_i$ into a spatially dependent component $f_x(\mathbf{s}_i)$ and residuals $r_{x_i}$: $ x_i = f_x(\mathbf{s}_i) + r_{x_i}$.
The model then replaces $X_i$ with $r_{x_i}$ in the spatial analysis model, resulting in:
$ Y_i = \beta r_{x_i} + f^+(\mathbf{s}_i) + \epsilon_i$,
where $f^+(\mathbf{s}_i)$ is also assumed to be a smooth surface that combines the spatial effects $f(\mathbf{s}_i)$ and $\beta f_x(\mathbf{s}_i)$.
This adjustment reduces the collinearity between $X_i$ and $f(\mathbf{s}_i)$, allowing for an unbiased estimation of $\beta$.
The method is straightforward to implement using existing regression tools like thin plate splines.
Another advantage of spatial+ is that all spatial information is retained in the model.
Many subsequent models developed from the Spatial+.

The Geoadditive Structural Equation Model (gSEM), introduced by \citet{2018_ThandenKneib_JASA}, is a direct predecessor to Spatial+ designed to rectify bias from unmeasured spatial confounding. When a shared spatial process affects both a covariate $\bX$ and the response $\bY$, standard geoadditive models may misattribute spatial variation to the covariate. gSEM addresses this by separately estimating spatial components for $\bX$ and $\bY$ within a structural equation framework, thereby isolating the direct covariate effect \(\beta_x\) from indirect spatial pathways.

The model is specified with two equations: \(X_i = \sum_k z_{ki}\gamma_{1k} + \varepsilon_{1i}\) and \(Y_i = X_i\beta_x + \sum_k z_{ki}\gamma_{2k} + \varepsilon_{2i}\), where \(\gamma_{1k}\) and \(\gamma_{2k}\) represent distinct spatial effects.
By explicitly modeling the separate spatial dependencies, gSEM provides a foundational approach for bias adjustment that Spatial+ later refines and simplifies.

\citet{2023_Marques_preprint} developed the Bayesian Spatial+ based on a joint model perspective.
The authors argue that adjusting for confounding in two stages using frequentist estimators neglects uncertainty propagation between stages.
They integrate the two stages into a joint Bayesian framework, allowing uncertainty propagation and direct parameter inference.
They also introduce joint priors to restrict spatial smoothness, ensuring unobserved effects do not operate at finer spatial scales than observed covariates.
%

In the original formulation \citep{2022_Dupont_Biometrics} there is no indication on how to conduct proper uncertainty quantification, but the Bayesian formulation makes the task straightforward.
\citet{2024_Urdangarin_SS} proposed a simplified version of Spatial+ for areal data, which instead of fitting a thin plate spline, filters the covariate $\bX$ by removing low-frequency eigenvalues.
They also extend the method to a multivariate response scenario. 
\citet{2024_Dupont_JABES} extend their method to non-linear covariate effects. 
They achieve this by using generalized additive models (GAMs, \cite{1986_Hastie_IMS}) to model the spatial effect, decomposing each element of the basis used into a spatial and residual components, and using the residuals as a new basis.

\subsection{Spectral Adjustment}

Motivated by the idea that spatial confounding can be scale-dependent, \citet{2022_Guan_Biometrika} recast the problem in the spectral domain and develop an adjustment that targets confounding at specific spatial frequencies.
They assume data coming from a continuous spatial domain and the same DGM as in Equation \eqref{Spatial_Generating_Model}.
Both processes $\mathbf{X}(s)$ and $\mathbf{Z}(s)$ have zero mean and are stationary, and thus have spectral representations $ \mathbf{X}(s) = \int e^{i \omega^\top s} \mathcal{X}(\omega) d\omega, \; \mathbf{z}(s) = \int e^{i \omega^\top s} \mathcal{Z}(\omega) d\omega$, 
where $\omega \in \mathbb{R}^2$ is a frequency.
The spectral processes $\mathcal{X}(\omega)$ and $\mathcal{Z}(\omega)$ are Gaussian with $\mathbb{E}(\mathcal{X}(\omega)) = \mathbb{E}(\mathcal{Z}(\omega)) = 0$ and are independent across frequencies, so that for any $\omega \neq \omega'$, $\text{Cov}\{\mathcal{Z}(\omega), \mathcal{Z}(\omega')\} = \text{Cov}\{\mathcal{X}(\omega), \mathcal{X}(\omega')\} = \text{Cov}\{\mathcal{X}(\omega), \mathcal{Z}(\omega')\} = 0$.
At the same frequency, the covariance has joint form
\begin{align*}
\text{Cov} \begin{pmatrix} \mathcal{X}(\omega) \\ \mathcal{Z}(\omega) \end{pmatrix} = 
\begin{pmatrix}
\sigma_x^2 f_x(\omega) & \rho \sigma_x \sigma_z f_{xz}(\omega) \\
\rho \sigma_x \sigma_z f_{xz}(\omega) & \sigma_z^2 f_z(\omega)
\end{pmatrix},
\end{align*}
where $\sigma_x^2$ and $\sigma_z^2$ are variance parameters, $f_x(\omega) > 0$ and $f_z(\omega) > 0$ are spectral densities that determine the marginal spatial correlation of $x(s)$ and $z(s)$, respectively, and the cross-spectral density $f_{xz}(\omega)$ determines the dependence between the spectral processes at different frequencies.
The conditional distribution of $\mathcal{Y}(\omega)$ given $\mathcal{X}(\omega)$, marginalizing over $\mathcal{Z}(\omega)$, is  $
\mathcal{Y}(\omega) \mid \mathcal{X}(\omega) \overset{\text{indep}}{\sim} \mathcal{N} \left( \beta_x \mathcal{X}(\omega) + \beta_z \alpha(\omega) \mathcal{X}(\omega), \tau^2(\omega) + \sigma^2 \right)$, where $
\alpha(\omega) = \frac{\rho \sigma_z f_{xz}(\omega)}{\sigma_x f_x(\omega)} = \frac{\sigma_z \sqrt{f_z(\omega)}}{\sigma_x \sqrt{f_X(\omega)}} \gamma(\omega), \;
\tau^2(\omega) = \beta_z^2 \sigma_z^2 f_z(\omega) \left[ 1 - \rho^2 \frac{f_{xz}(\omega)^2}{f_x(\omega) f_z(\omega)} \right]$.

The regression coefficient for $\mathcal{X}(\omega)$ is $\beta(\omega) = \beta_x + \beta_z \alpha(\omega) \neq \beta_x$.
The additional term $\widehat{\mathcal{Z}}(\omega) = \mathbb{E}[\mathcal{Z}(\omega) \mid \mathcal{X}(\omega)] = \alpha(\omega) \mathcal{X}(\omega)$ is the result of attributing the effect of the unmeasured confounder to the response to the treatment variable, which could induce bias in estimating $\beta_x$.

It is then clear that assumptions about the nature of $\alpha(\omega)$ are necessary to correct for confounding.
The authors propose two approaches for the correct identification of the effect $\beta_x$:
unconfoundedness at high frequencies, that is, if we assume that $\alpha(\omega) \to 0$ for large $\|\omega\|$, then $\mathbb{E}(\mathcal{Y}(\omega) \mid \mathcal{X}(\omega)) \approx \beta_x \mathcal{X}(\omega)$, and thus $\beta_x$ is identified;
a parsimonious and parametric model with constraints on the parameters.
This in turn implies that the correlations between $\mathbf{X}$ and $\mathbf{Z}$ are constant between all frequencies.
It is possible to verify that with this parsimonious specification, all parameters are identifiable.

Returning to the spatial domain, the adjustment for confounding is then done by adjusting the following model
\begin{align*}
\mathbf{Y}(s) \mid \mathbf{X}(s), s \in \mathcal{D} &= \beta_0 + \beta_x \mathbf{X}(s) + \beta_z \widehat{\mathbf{Z}}(s) + \delta(s), \\
\widehat{\mathbf{Z}}(s) &= \int \exp(i \omega^T s) \widehat{\mathcal{Z}}(\omega) d\omega = \int \exp(i \omega^T s) \alpha(\omega) \mathcal{X}(\omega) d\omega.
\end{align*}

Using $\hat{\mathbf{Z}}$ as a covariate can alleviate confounding in some smoothing configurations, as described by the authors. 
To produce $\hat{\mathbf{Z}}$, a parsimonious bivariate Matérn model can be assigned to $\mathbf{X}$ and $\mathbf{Z}$.
The other proposal of the authors is to fit a semi-parametric model by using B-splines to model $\alpha(\omega)$.

In a more recent contribution, \citet{2025_Prim_spectralconfounderadjustmentspatial} extend spectral methods to handle multiple exposures and multiple outcomes simultaneously.
They model multiscale effects using a three-way tensor (exposures x outcomes x spatial scales), applying canonical polyadic decomposition to the tensor to induce low-rank structures and enable parameter sharing across exposures and outcomes.
They implement it by Bayesian tensor regression using horseshoe priors \citep{2010_CarvalhoPolsonScott_Horseshoe} to promote sparsity and prevent overfitting.


\subsection{Correlating Gaussian Random Fields}

\citet{2022_Marques_Env} also construct a method inspired by the joint mechanism generating observed and unobserved quantities.
They propose a joint Gaussian distribution for $\bX$ and $\bZ$.
The authors hypothesize that explicitly correlating Gaussian random fields for $\bX$ and $\bZ$ in spatial regression models can reduce bias in regression coefficient estimates caused by spatial confounding. 
Specifically, by jointly modeling the spatial random effect and the covariates using a multivariate Gaussian random field, the proposed model aims to better account for the spatial dependencies and reduce confounding.
Let \(\mathbf{Z} \sim \mathcal{N}(\mathbf{0}, \boldsymbol{\Sigma}_{z})\).
They assume that \(\mathbf{Z}(s)\) and \(\mathbf{X}(s)\) are jointly Gaussian distributed such that
\[
\begin{pmatrix} 
\boldsymbol{\gamma} \\ 
\mathbf{Z} 
\end{pmatrix} 
\sim \mathcal{N} \left(
\begin{pmatrix} 
\mathbf{0} \\ 
\boldsymbol{\mu}_z 
\end{pmatrix}, 
\begin{pmatrix} 
\boldsymbol{\Sigma}_{\gamma} & \rho \boldsymbol{\Sigma}_{\gamma}^{1/2} (\boldsymbol{\Sigma}_{z}^{1/2})^T \\ 
\rho \boldsymbol{\Sigma}_{z}^{1/2} (\boldsymbol{\Sigma}_{\gamma}^{1/2})^T & \boldsymbol{\Sigma}_z 
\end{pmatrix} 
\right).
\]
In this setup, all conditional distributions can be easily calculated, allowing us to calculate the distribution of the spatial effect conditional on the observations.

For the parameter $\rho$, a Penalized Complexity (PC) priors are used.
PC priors to the correlation parameter can provide a computationally efficient way to shrink toward a non-confounded base model.
That is, the model allows non-zero $\rho$ only if the data strongly supports it, ensuring that spatial confounding is addressed dynamically.
 The prior takes the form 
 $$p(\rho) = \frac{c - 1}{2} \left( \frac{1}{1 - \rho} - \frac{1}{1 + (c - 1)\rho} \right) 
\frac{\lambda}{\sqrt{-lR(\rho)}} \exp(-\lambda \sqrt{-lR(\rho)}),$$
where \( lR(\rho) = \log(R(\rho)) \) and \( R(\rho) = (1 + (c - 1)\rho)(1 - \rho)^{c-1} \).
The decay-rate \(\lambda\) can be chosen by sampling penalized complexity (PC) priors~\citep{Simpson2017} from various values of \(\lambda\) and choosing a \(\lambda\) satisfying $\text{Prob}(|\rho| > U) = \alpha$.

The method generalizes to cases with multiple spatially confounded covariates, improving interpretability and model estimation.
\citet{2022_Marques_Env} also includes the case of multiple covariates using PCA for dimensionality reduction.

\section{An analytical framework for confounding}
\label{sec:analytical_frame}

To resolve the long-standing confusion surrounding spatial confounding, \citet{2023_Dupont_preprint} proposed a unifying theoretical framework.
Unlike previous works that relied on empirical observations and simulations, this framework provides explicit analytical expressions for bias, identifying spatial smoothing as the primary mechanism that reintroduces confounding. 

We consider the linear spatial model as the data generating mechanism, analyzed via:
\begin{eqnarray}
\label{eq:analysis}
    \bY &=& \beta \bX + \mathbf{B}_{sp} \boldsymbol{\beta}_{sp} + \boldsymbol{\varepsilon}, \quad \boldsymbol{\beta}_{sp} \sim \mathcal{N} (\mathbf{0}, \lambda^{-1} \mathbf{S}^{-})
\end{eqnarray}
where the unobserved confounder $\bZ$ is approximated by the spatial effect $\gamma(s) = \mathbf{B}_{sp} \boldsymbol{\beta}_{sp}$. The matrix $\mathbf{S}$ is the penalty matrix with eigenvalues $0 = \alpha_1 \leq \dots \leq \alpha_p$. 

The core of the confounding mechanism lies in the properties of the spatial precision matrix $\mathbf{\Sigma}^{-1}$.
Using the eigen-decomposition of $\mathbf{\Sigma}^{-1}$, \citet{2023_Dupont_preprint} derive weights $w_i = \lambda \alpha_i / (\sigma^{-2} + \lambda \alpha_i)$, which represent the degree of smoothing applied to different spatial frequencies.
In what follows we present a few technical results taken from \cite{2023_Dupont_preprint} -- the interested reader is referred to the original paper for the proofs.

\begin{lemma}
\label{lem:eigen}
\textit{Let} $\alpha_1 \leq \dots \leq \alpha_p$ \textit{be the eigenvalues of the penalty matrix} $\mathbf{S}$ \textit{and} $\lambda > 0$ \textit{the smoothing parameter. Then the eigenvalues of the precision matrix} $\mathbf{\Sigma}^{-1}$ \textit{are given by} $\{\sigma^{-2}, \sigma^{-2} w_1, \dots, \sigma^{-2} w_p\}$, \textit{where} $w_i = \lambda \alpha_i / (\sigma^{-2} + \lambda \alpha_i)$ \textit{for} $i = 1, \dots, p$.
\end{lemma}

The columns of the associated eigenvector matrix $\mathbf{U}$ form an orthonormal basis.
The first $n - p$ eigenvectors, $\mathbf{U}_{ns}$, span the non-spatial subspace, while the remaining $p$ eigenvectors, $\mathbf{U}_{sp}$, correspond to spatial components. Since $w_i \in [0,1]$, spatial eigenvectors with weights close to 1 are called \textit{high-frequency}, while those with low weights are \textit{low-frequency}.

\begin{proposition}
\label{prop:bias_covariate}
    \textit{The bias of the estimated covariate effect} $\hat{\beta}$ \textit{in model \eqref{eq:analysis} is governed by the projection of the exposure and the confounder onto the precision metric:}
\[
\mathbb{E}(\hat{\beta}) - \beta = \frac{\langle \mathbf{X}, \mathbf{Z} \rangle_{\mathbf{\Sigma}^{-1}}}{\|\mathbf{X}\|^2_{\mathbf{\Sigma}^{-1}}}.
\]
\end{proposition}

By projecting $\mathbf{X}$ and $\mathbf{Z}$ onto the eigenbasis $\mathbf{U}$, where $\boldsymbol{\xi}^x$ and $\boldsymbol{\xi}^z$ are the respective coordinates, the bias can be expressed as:
\begin{equation}
\label{eq:bias_prop}
\text{Bias}(\hat{\beta}) = \frac{\sum_{i=1}^{p} \xi^{x}_{sp,i} \xi^{z}_{sp,i} w_i}{\sum_{i=1}^{n-p} (\xi^{x}_{ns,i})^2 + \sum_{i=1}^{p} (\xi^{x}_{sp,i})^2 w_i}.
\end{equation}

This expression "demystifies" several key phenomena.
First, it identifies smoothing as the cause of the bias; if no smoothing is applied, then  and the bias disappears, revealing that bias arises precisely because we "penalize" the spatial effect to prevent overfitting (a bias-variance trade-off), which prevents the model from fully "absorbing" the confounder.
Second, it highlights frequency dependence, where confounding at high frequencies causes the most severe bias because the model heavily penalizes these components, forcing the spatial signal into the  estimate, whereas low-frequency confounding is largely absorbed by the spatial random effect, leaving  relatively unbiased.

This framework also clarifies why Restricted Spatial Regression (RSR) fails to solve the problem.
RSR assumes that the spatial effect and the covariate should be independent. However, as \citet{2023_Dupont_preprint} argue, if $\bZ$ is a true confounder, it is by definition correlated with $\bX$. Thus, the "independence" enforced by RSR effectively ignores the confounding, leading to estimates that are identical to biased non-spatial models.

\subsection{Consistency of common spatial estimators}

While the framework above explains why spatial models are biased in practice, recent work by \cite{2024_Gilbert_Biometrika} and \cite{2025_Datta_preprint} investigates whether this bias vanishes asymptotically ($n \to \infty$).
\cite{2024_Gilbert_Biometrika} demonstrate that GLS remains consistent despite model misspecification, provided the exposure contains non-spatial variation. This consistency is achieved because the GLS transformation acts as a \textit{prewhitening} filter.

However, \cite{2025_Datta_preprint} reveal a "spectral threshold" for this consistency: if the exposure $\mathbf{X}$ is smoother than the confounder $\bZ$ by a factor of more than $d/2$ in a $d$-dimensional space, the slope is no longer consistently estimable.
This aligns with the frequency-based analysis in \citet{2023_Dupont_preprint}: when the "signal" of $\bX$ is buried in the low frequencies where smoothing is weakest, the model cannot distinguish the covariate effect from the spatial confounding, resulting in persistent bias.


\section{A Causal Inference perspective}
\label{sec:causal}

While confounding lies at the very heart of causal inference, the term was used in the spatial statistics literature for over a decade before receiving a formal, strictly causal treatment.
Recently, a growing body of research has sought to provide a more robust meaning to spatial confounding by explicitly drawing upon methodologies and identification strategies from causal inference theory \citep{2018_ThandenKneib_JASA, 2019_Papadogeorgou_Biostatistics, 2020_Schnell_AAS, 2021_Reich_ISR, 2024_Gilbert_preprint}.

Through a causal lens, confounding is defined by the data-generating process: it is omitted-variable bias from a (possibly unobserved) common cause of exposure and outcome.
In spatial settings, that omitted variable typically has spatial structure, so “spatial confounding” is treated here as bias from an unmeasured spatially structured confounder, and the goal becomes identifying what assumptions and adjustments are sufficient to recover a causal effect.

To integrate spatial statistics with causal inference, we adopt the potential outcomes framework \citep{1974_Rubin_JEP}.
Let $\mathbf{Y}(x)$ denote the potential outcome under exposure level $x$. Identifying causal targets—such as the Average Treatment Effect (ATE), $\mathbb{E}[\mathbf{Y}(1) - \mathbf{Y}(0)]$, or the dose-response curve, $\mathbb{E}[\mathbf{Y}(x)]$—from observational data requires three standard assumptions:
(i) \textbf{Consistency}: $\mathbf{Y} = \mathbf{Y}(\mathbf{x})$;
(ii) \textbf{Positivity}: $f(x|\mathbf{\mathbf{C}}) > 0$ for all $x$ in the support of $\mathbf{X}$; and
(iii) \textbf{Exchangeability}: $Y(x) \perp X \mid \mathbf{C}$.
Under these conditions, the causal effect is identifiable via the propensity score $e(\mathbf{C}) = f(\mathbf{X} \mid \mathbf{C})$, typically estimated using Inverse Probability Weighting (IPW, \cite{2001_hirano_IPW}) or Doubly Robust methods \citep{2005_bang_doubly}.

Though unobserved, unmeasured confounders $\mathbf{U}$ are typically spatially structured, enabling the use of location $\mathbf{S}$ as a proxy.
 \citet{2024_Gilbert_preprint} formalizes this strategy through two key adaptations to standard causal identification assumptions.
The standard assumption of exchangeability ($\mathbf{Y}(x) \perp \mathbf{X} \mid \mathbf{C}$) requires us to condition on all common causes.
If $\mathbf{U}$ is unobserved, exchangeability is violated.
To recover identification, we must assume that the spatial location $\mathbf{S}$ contains all the information necessary to represent $\mathbf{U}$.
This leads to the \textbf{Measurability Assumption}: $\mathbf{U} = g(\mathbf{U})$, where $g(\cdot)$ is a measurable function.
Under this assumption, conditioning on the spatial coordinates $\mathbf{S}$ is equivalent to conditioning on the confounder itself. Consequently, if we assume that potential outcomes are independent of exposure given $U$, then they are also independent given $\mathbf{S}: \mathbf{Y}(x) \perp \mathbf{X} \mid \mathbf{S}$.
This justifies the common practice in spatial statistics of including a spatial random effect or a smooth surface $\gamma(\mathbf{S})$ in a regression.
From a causal perspective, this surface is not just a way to account for residuals; it is an attempt to model the function $g(\mathbf{X})$ to satisfy exchangeability.

While the measurability assumption allows us to treat $\mathbf{S}$ as a proxy for the confounder, identification still requires an overlap/positivity condition relative to the variable we condition on.
If $\mathbf{X}$ were also a purely spatial process, meaning $\mathbf{X} = h(\mathbf{S})$, then within any fixed s there is essentially no remaining exposure variation, so causal contrasts are unidentified without extrapolation beyond the observed $(\mathbf{X},\mathbf{S})$ support.
\citet{2024_Gilbert_Biometrika} therefore require that positivity hold with respect to spatial location, for example by assuming directly that $(x,s)$ occurs whenever $x \in supp(\mathbf{X})$ and $s\in supp(S)$, or indirectly by combining standard positivity conditional on $\mathbf{U}$ with the assumption that $\mathbf{S}$ is associated with $\mathbf{X}$ only through $\mathbf{U}$, which implies positivity conditional on $\mathbf{S}$.
In effect, this aligns with the requirement identified by \citet{2010_Paciorek_StatSci} that the exposure must vary at a finer spatial scale than the confounder.
If $\mathbf{X}$ is "rougher" than the confounder $\mathbf{U}$, then in a small geographic neighborhood where $\mathbf{U}$ is approximately constant, we still observe a range of values for $\mathbf{X}$.
This local variation provides the contrast necessary to identify the causal effect.

\subsection{Propensity Score Adjustment}
\label{sec:causal_propscore}

\citet{2019_Davis_SMMR} develops a doubly robust estimator for the average treatment effect that accounts for spatial information.
The method is based on the estimator for the CATE: $\Delta = \mathbb{E}[Y_1 - Y_0|T,X]$, $    \hat{\Delta} = n^{-1} \sum_{i=1}^{n}\left[\frac{T_i Y_i}{\hat{e}_i} - \frac{(1 - T_i) Y_i}{1-\hat{e}_i}\right]$,
where $\hat{e}_i$ is an estimator for the propensity score, usually a logistic regression.
Note that in this first estimator IPW is being applied to reweigh observations.
To provide a doubly-robust estimator \citet{2019_Davis_SMMR} use the methods by \citet{1994_Robins_JASA} to modify the estimator above to $\hat{\Delta}^{DR} = \frac{1}{n} \sum_{i=1}^n \hat{\Delta}_i$ such that $\hat{\Delta}_i = \left[\frac{T_i Y_i}{\hat{e}_i} - \frac{(T_i - \hat{e}_i)\hat{Y}_{i1}}{\hat{e}_i}\right] 
- \left[\frac{(1 - T_i)Y_i}{1 - \hat{e}_i} + \frac{(T_i - \hat{e}_i)\hat{Y}_{i0}}{1 - \hat{e}_i}\right]$.

Here $\hat Y_{i1}$ and $\hat Y_{i0}$ are predicted outcomes from an outcome regression of $Y$ on $(X,T)$, where $\hat Y_{i1}$ includes the effect of $T$ (equivalently, sets $T=1$) and $\hat Y_{i0}$ excludes it (equivalently, sets $T=0$).
In order to accommodate spatial dependency in this procedure, \citet{2019_Davis_SMMR} proposes to substitute the outcome regressions and the logistic regression to adjust the propensity score to a spatial regression, in particular an ICAR model.


\subsection{Distance-adjusted propensity score matching}
\label{sec:causal_daps}

\citet{2019_Papadogeorgou_Biostatistics} develop a propensity score matching method applied to spatially indexed data.
Propensity score matching \citep[PSM,][]{1983_RosenbaumRubin_Biometrika} matches observations with similar propensity scores.
This helps to construct an artificial control group by matching each treated unit with a non-treated unit of similar characteristics, thus resampling a RCT.
Matches can be made by nearest neighbor, caliper matching (a propensity score threshold is chosen), exact matching, and many others.

Distance-adjusted propensity score matching (DAPS) augments confounding adjustment \textit{via} PSM by incorporating spatial information as a proxy for unobserved spatial variables.
DAPS combines propensity score estimates and relative distances to define $\operatorname{DAPS}_{ij} = w \cdot |PS_i - PS_j| + (1 - w) \cdot Dist_{ij}$, where $w \in [0, 1]$, and $PS_i$ and $PS_j$ are propensity score estimates from modeling treatment conditional on observed confounders, and \(Dist_{ij}\) is a distance measure for observations $i$ and $j$.
The authors also develop a data-driven method to estimate the parameter $w$.
The choice of $Dist$ is made by normalizing the distances between pairs of points, so $Dist$ falls into $[0,1]$.

\subsection{Double machine learning for spatial confounding}
\label{sec:causal_dml}

Recently, \citet{2024_Wiecha_DML} developed double machine learning (DML) methods to address spatial confounding. DML, originally proposed by \citet{2018_Chernozhukov_DML}, is designed to address high-dimensional confounding and model selection bias by using two-stage semiparametric estimation, ensuring root-n consistency and asymptotic normality of the estimated treatment effect.
They accomplish this with a similar method to \cite{2018_ThandenKneib_JASA}. 
The first stage comprises of adjusting a GP with Matérn covariance to both exposure and outcome.
Then, a final stage is adjusted regressing the outcome on the exposure. 
They manage to accomplish this while ensuring all desirable theoretical properties for DML estimators, including root-n-consistency.

In their notation, the method is developed for the partially linear spatial model $Y_i = A_i^\top \beta_0 + Z_i^\top \theta_0 + g_0(S_i) + U_i$ and $A_{ij} = Z_i^\top \theta_j + m_{0j}(S_i) + V_{ij}$, $j=1,\dots,\ell$, where $S_i\in\mathbb{R}^d$ is the location, $g_0(\cdot)$ and $\{m_{0j}(\cdot)\}$ are unknown spatial trends, and $\beta_0$ is the parameter of interest. To estimate the spatial trends, \citet{2024_Wiecha_DML} use Gaussian process regression with a Mat\'ern correlation function~\eqref{eq:Matern_spec}.
Let $\widehat{g}_0(S_i)$ and $\widehat{m}_{0j}(S_i)$ denote (cross-fitted) GP/Kriging predictors of $g_0(S_i)$ and $m_{0j}(S_i)$, respectively, and define residualized regressors $\widehat{V}_{ij} = A_{ij} - \bigl(Z_i^\top \widehat{\theta}_j + \widehat{m}_{0j}(S_i)\bigr)$, $j=1,\dots,\ell$, $\widehat{U}_i = Y_i - \mathbf{A}_i^\top \widehat{\boldsymbol{\beta}}_{\mathrm{DSR}} - \mathbf{Z}_i^\top \widehat{\boldsymbol{\theta}}_0 - \widehat{g}_0(S_i)$.
Stacking $\widehat{V}_{ij}$ into $\widehat{\mathbf{V}}\in\mathbb{R}^{n\times \ell}$, the Double Spatial Regression (DSR) / DML estimator is $\widehat{\boldsymbol{\beta}}_{\mathrm{DSR}}
=
(\widehat{\mathbf{V}}^\top \mathbf{A})^{-1}\widehat{\mathbf{V}}^\top\!\left(\mathbf{Y} - \mathbf{Z}^\top \widehat{\boldsymbol{\theta}}_0 - \widehat{g}_0(S)\right)$, with a closed-form heteroskedasticity-robust variance estimator.


\section{Results on real datasets}
\label{sec:case_studies}

We now illustrate the behavior of competing methods on the real datasets listed above.  
For each case study we focus on the coefficient(s) of primary scientific interest, and address three questions: (i) how the inclusion of spatial random effects changes the magnitude and uncertainty of the estimate compared with a non-spatial model; (ii) whether methods designed to mitigate spatial confounding (RSR, Spatial$+$, spectral adjustment, SPOCK) behave in line with their intended purpose; and (iii) whether any method displays anomalous behavior relative to the others.  
All results are reported as point estimates with corresponding 95\% intervals: for frequentist methods we display the estimate and its 95\% confidence interval, and for Bayesian methods we display the posterior mean and the 95\% credible interval.
All analyses are fully reproducible using the code available at our GitHub repository: \url{https://github.com/isaquepim/spatial-confounding}.


\subsection{Scotland lip cancer}

\begin{figure}[ht]
  \centering
  \includegraphics[width=0.48\linewidth, height=0.25\textheight]{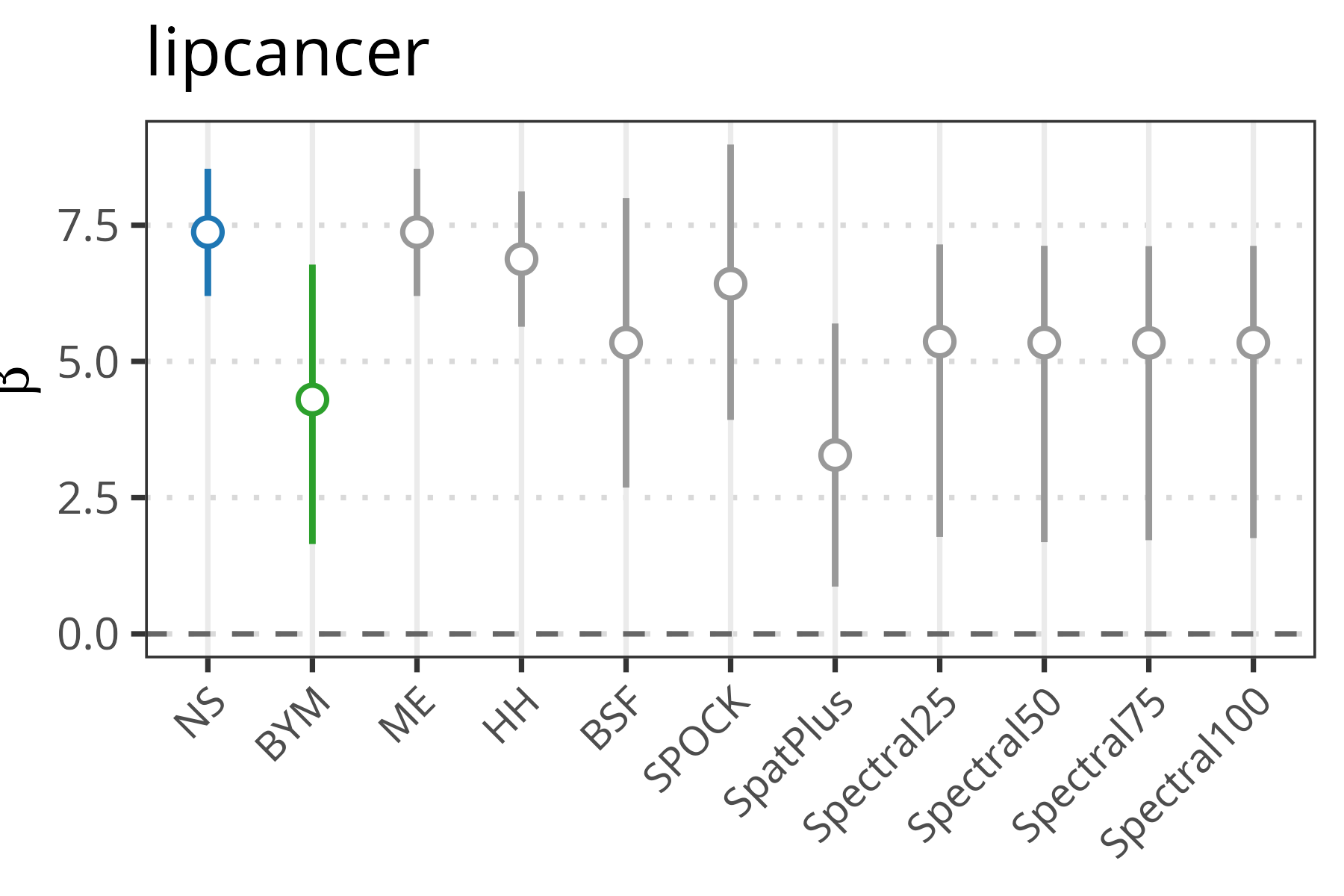}\hfill
  \includegraphics[width=0.48\linewidth, height=0.25\textheight]{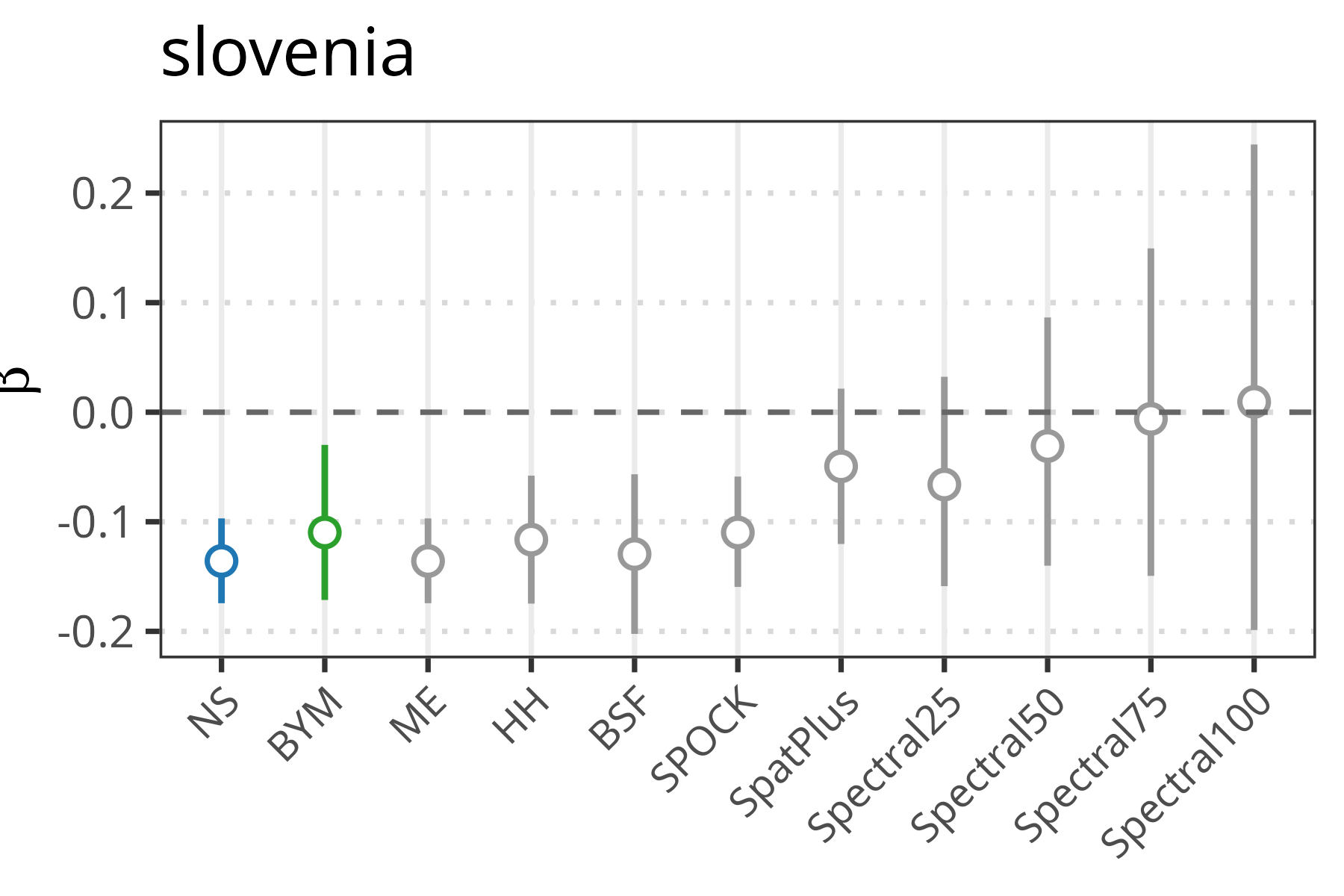}
  \caption{Estimated effects across all methods. Left: Scotland lip cancer (AFF). Right: Slovenia stomach cancer (socio-economic status). frequentist methods are shown as point estimates with 95\% confidence intervals and Bayesian methods as posterior means with 95\% credible intervals.}
  \label{fig:scotland_slovenia}
\end{figure}

We begin our exploration with a very famous dataset from spatial statistics, namely the Scotland lip cancer dataset~\citep{Clayton1987}.
For the Scotland lip cancer data, the outcome is county-level lip cancer incidence and the main covariate is the percentage of the population employed in agriculture, fishing and forestry (AFF) (Figure~\ref{fig:scotland_slovenia}, left panel).  
Including BYM spatial random effects reduces the estimated effect of AFF and increases its uncertainty compared with the non-spatial model, which is consistent with interpreting the spatial random effects as a regularizer.  
The RSR estimates are close to those from the non-spatial model while still accounting for residual spatial correlation, in line with the goal of recovering the marginal AFF effect.  
Spatial$+$ and the spectral adjustments (Spectral25 to Spectral100) yield point estimates that align more closely with the BYM estimates.  
No method produces an estimate that is clearly at odds with the others, and together they support a positive association between AFF and lip cancer risk, with some attenuation when spatial structure is explicitly modeled.

\subsubsection{Slovenia stomach cancer}

In the Slovenia stomach cancer dataset~\citep{Zadnik2006}, cancer incidence is regressed on socio-economic status (Figure~\ref{fig:scotland_slovenia}, right panel).  
The non-spatial and BYM models yield similar directions of effect, with the spatial model moderately attenuating the estimate, indicating that spatial structure plays a role but does not overturn the non-spatial conclusion.  

For this dataset, under spectral adjustment the low-frequency components capture most of the signal, whereas the high-frequency spectral coefficients are highly uncertain, reflecting limited residual information at small spatial scales.  
This is consistent with socio-economic status varying smoothly over space.  
Consequently, spectral estimates for the main covariate remain close to those from the traditional spatial models, and filtering out finer scales does not materially change the substantive inference.  

In contrast, the Spatial$+$ specification yields an estimated effect of socio-economic status that is strongly shrunk towards zero, effectively removing the association seen in the other models.  
The spectral fits show that the contribution of socio-economic status is concentrated in the smooth, low-frequency components and becomes negligible at higher frequencies, so aggressive filtering of these components can substantially weaken the estimated effect, providing an explanation for the Spatial$+$ result.

\subsection{Pennsylvania lung cancer}

\begin{figure}[ht]
  \centering
  \includegraphics[width=0.48\linewidth, height=0.25\textheight]{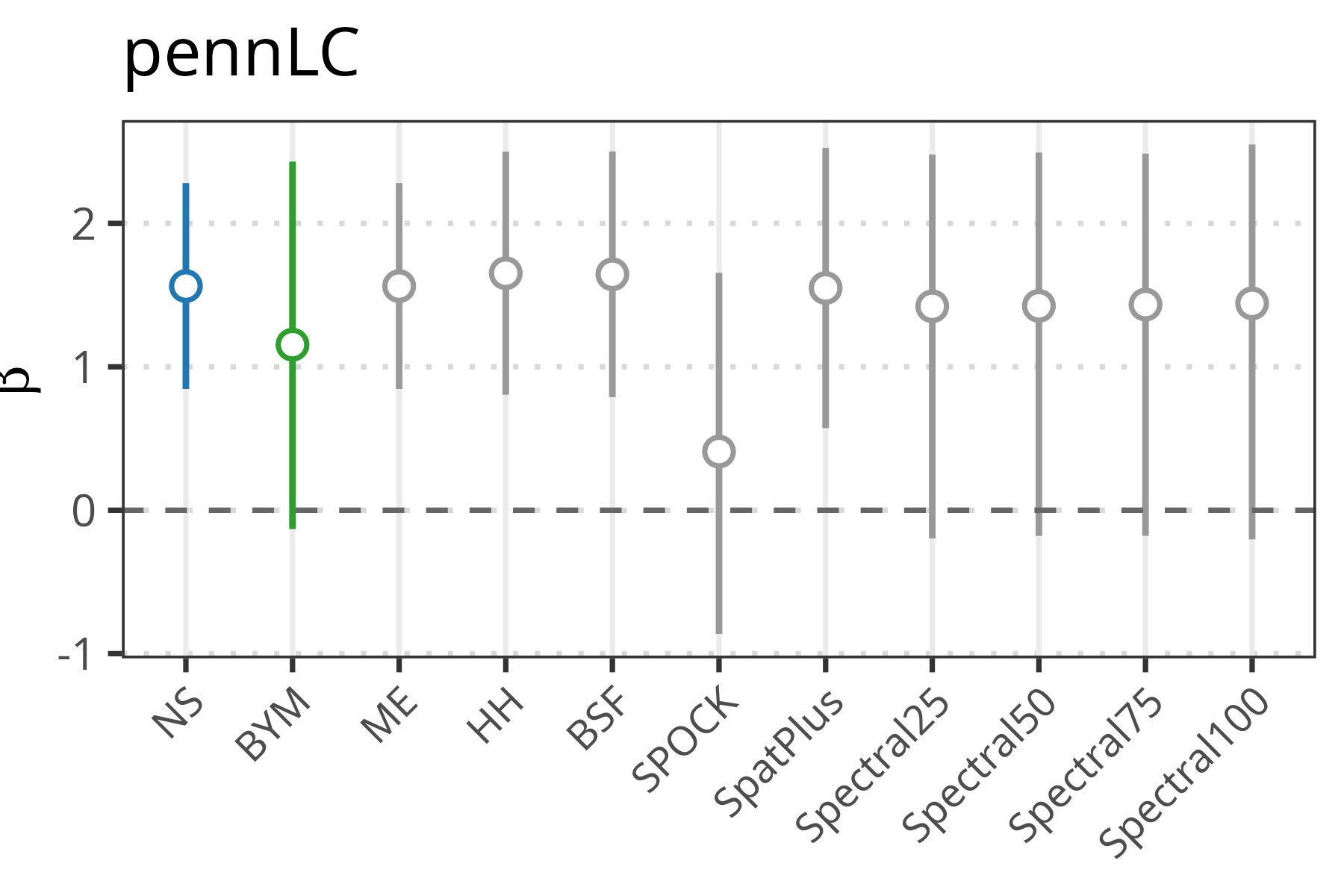}\hfill
  \includegraphics[width=0.48\linewidth, height=0.25\textheight]{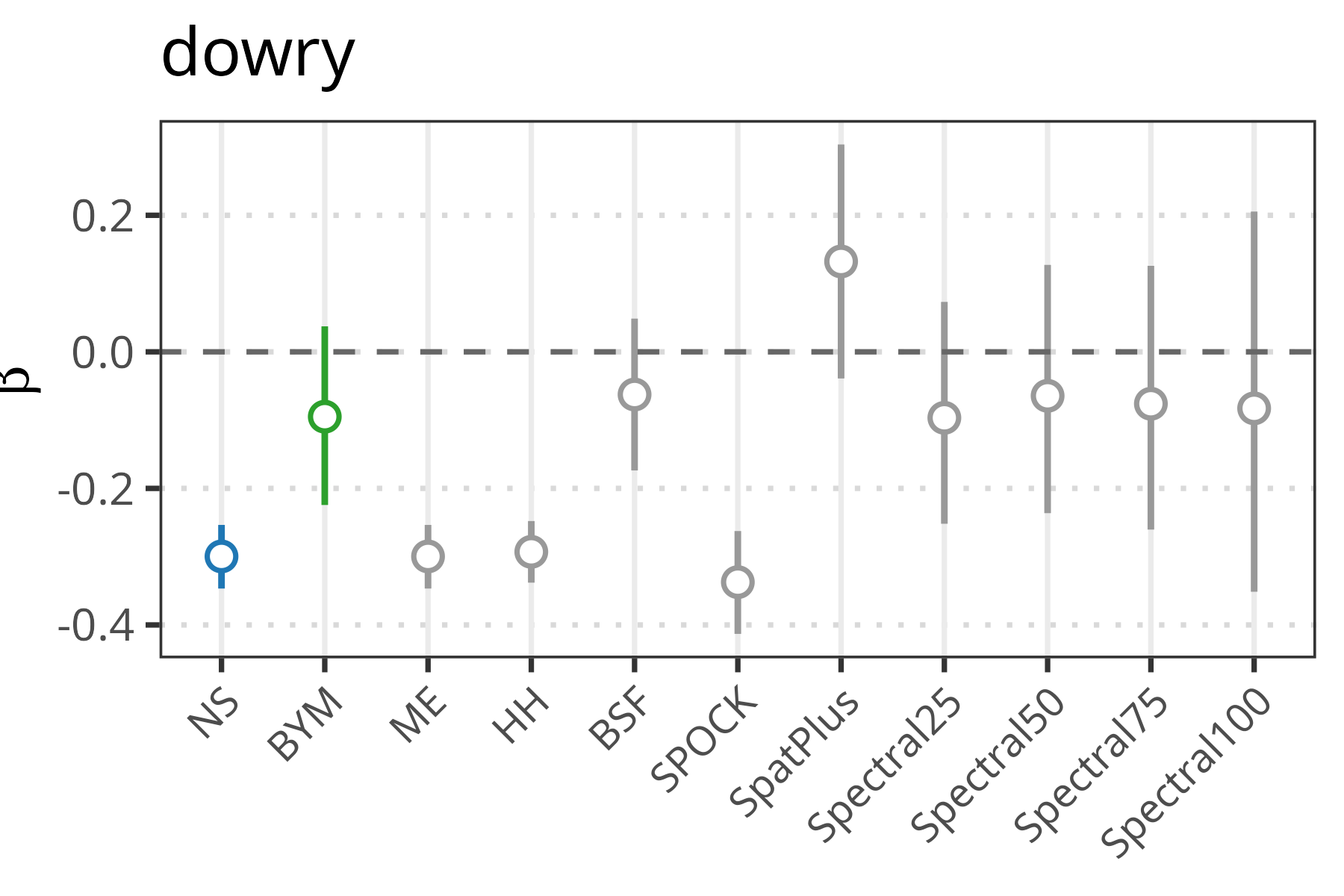}
  \caption{Estimated effects across all methods. Left: Pennsylvania lung cancer (smoking prevalence). Right: Dowry deaths in Uttar Pradesh (key socio-economic covariate). frequentist methods are shown as point estimates with 95\% confidence intervals and Bayesian methods as posterior means with 95\% credible intervals.}
  \label{fig:penn_dowry}
\end{figure}

For the analysis of lung cancer in Pennsylvania~\citep{SpatialEpi}, lung cancer incidence is regressed on county-level smoking prevalence (Figure~\ref{fig:penn_dowry}, left panel). 
The non-spatial model suggests a strong positive association: higher smoking prevalence is linked to higher lung cancer risk, with a 95\% interval that excludes zero.  
Under the BYM model, this effect is attenuated and its interval may include zero, illustrating the classic signature of spatial confounding where spatial random effects absorb part of the association between exposure and outcome.

The confounding-adjusted methods help clarify this discrepancy.  
In our analysis, RSR and Spatial$+$ yield estimates that are closer to the non-spatial coefficient, while still accounting for spatial structure, thereby recovering a positive and statistically important association between smoking and lung cancer.  
SPOCK, however, deviates from this pattern: its coefficient is noticeably different from that of the other methods, and its interval shows a different degree of uncertainty.  
This discrepancy suggests that, for this dataset, the SPOCK transformation behaves differently from the other confounding-adjusted approaches, and its result should be interpreted with caution.

\subsection{Dowry deaths in Uttar Pradesh}

The dowry deaths data from Uttar Pradesh~\citep{2023_Aritz_StatMod} provide an example of a social outcome with strong spatial structure, modeled as a function of socio-economic and demographic indicators (Figure~\ref{fig:penn_dowry}, right panel).  
Here, non-spatial regression suggests a clear association between a key socio-economic covariate and dowry-related mortality.  
Introducing BYM spatial effects reduces the magnitude of the association and widens the interval, so that the effect of this variable is no longer statistically distinguishable from zero.

The confounding-adjusted models further shrink the coefficient towards zero, effectively removing the effect of the covariate from the analysis and indicating that additional confounding mechanisms may be present, as also noted by \citet{2023_Urdagarin_RMC}.  
In contrast, the RSR specifications retain an effect that is similar in magnitude and significance to that obtained under the non-spatial model (NS), suggesting that they recover part of the marginal association while still accounting for spatial structure.

\subsection{Forestry data}

The forestry dataset from \citet{2022_Dupont_Biometrics} consists of spatially referenced measurements of tree growth, regressed on two key environmental covariates: tree age and May minimum temperature.  
Gaussian spatial models with splines capture smooth spatial variation in the response, while non-spatial regression treats observations as independent, and continuous RSR and Spatial$+$ aim to reduce potential spatial confounding.

\begin{figure}[t]
  \centering
  \includegraphics[width=0.48\linewidth]{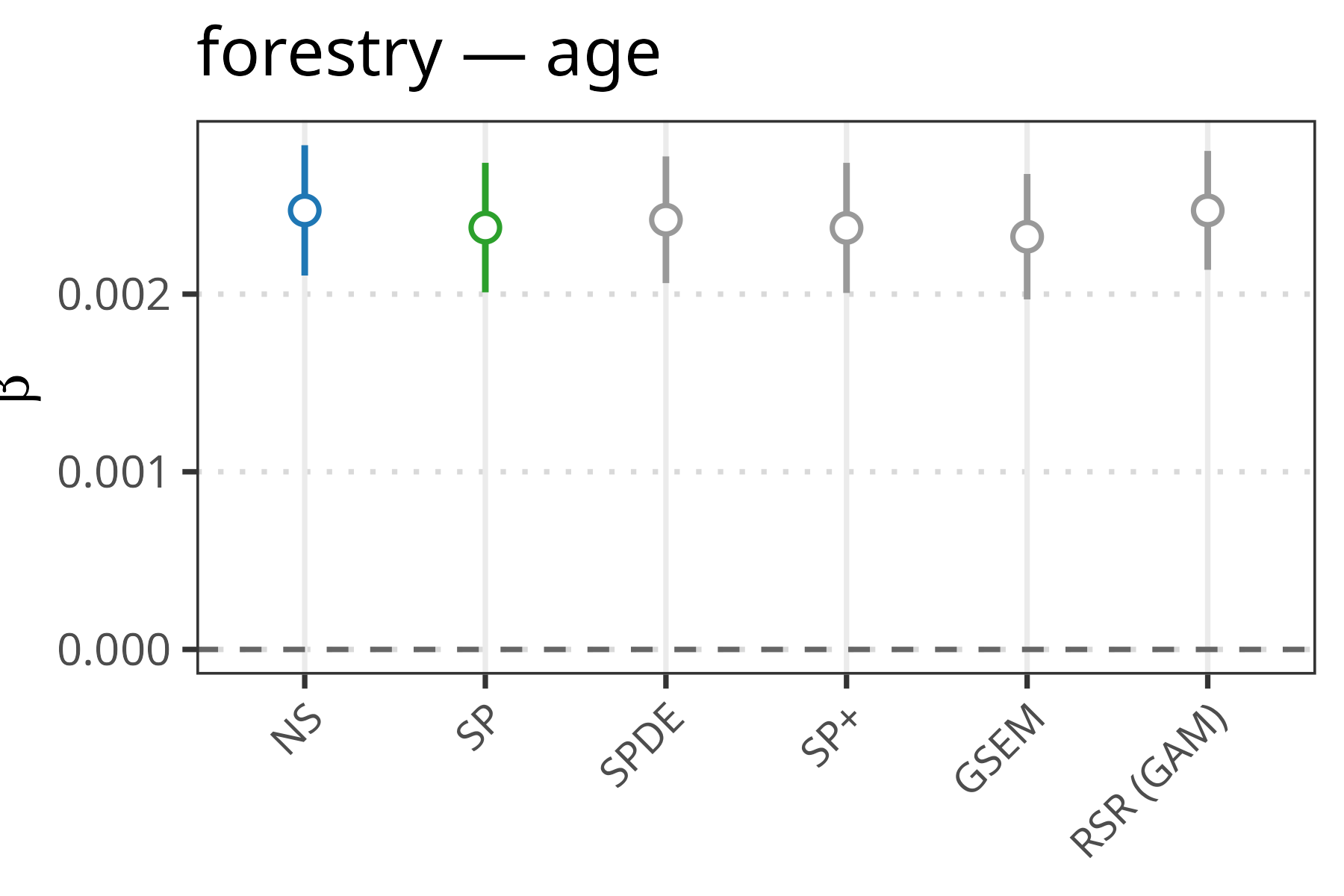}\hfill
  \includegraphics[width=0.48\linewidth]{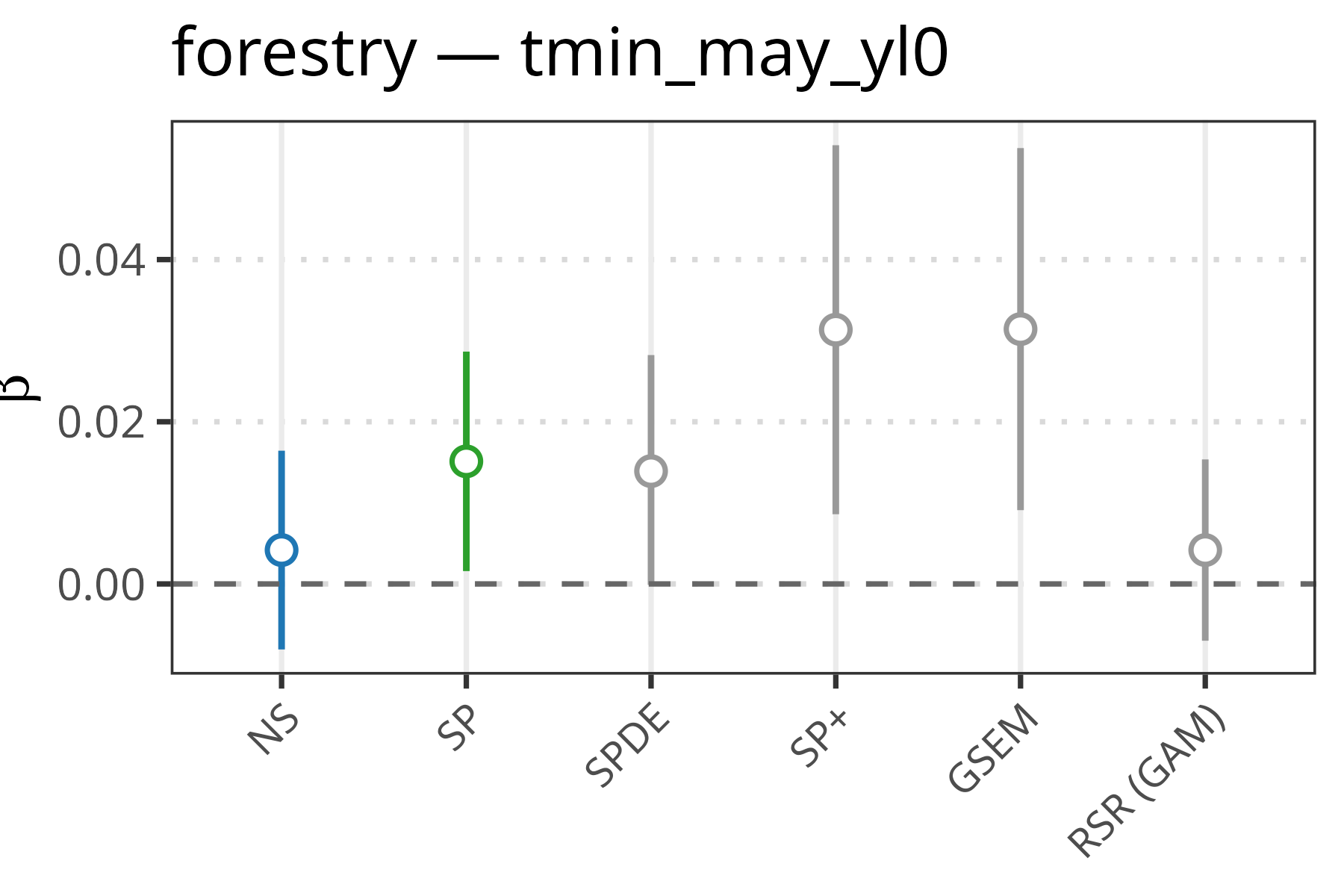}
  \caption{Forestry data. Estimated effects of (left) tree age and (right) May minimum temperature across all methods. Frequentist methods are shown as point estimates with 95\% confidence intervals and Bayesian methods as posterior means with 95\% credible intervals.}
  \label{fig:forestry}
\end{figure}

For tree age (left panel of Figure~\ref{fig:forestry}), all methods yield very similar estimates and overlapping 95\% intervals.  
The non-spatial, spatial spline, RSR and Spatial$+$ models agree both on the magnitude and the significance of the age effect, indicating that spatial confounding plays little role for this covariate and that its relevance for tree growth is robust to the choice of model.

For May minimum temperature (right panel of Figure~\ref{fig:forestry}), the picture is different.  
In the non-spatial and basic spatial spline models, the estimated effect is weaker and its interval is closer to including zero, suggesting a more uncertain association.  
After correcting for spatial confounding, particularly under RSR and Spatial$+$, the May minimum temperature coefficient becomes more clearly distinct from zero, indicating a meaningful association with growth once its correlation with the latent spatial field is accounted for.  

\subsection{Malaria in Gambia}

\begin{figure}[t]
  \centering
  \includegraphics[width=0.48\linewidth]{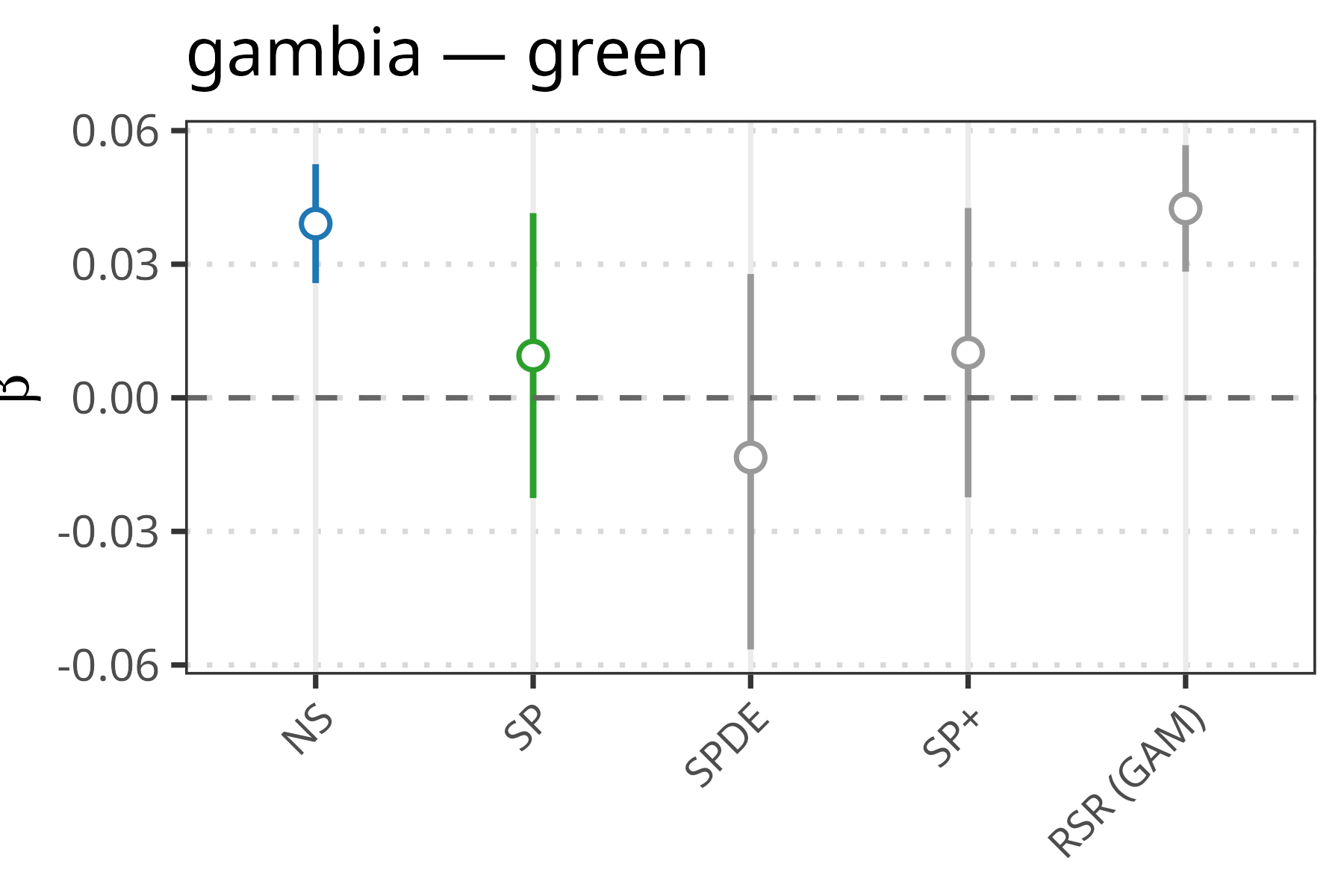}\hfill
  \includegraphics[width=0.48\linewidth]{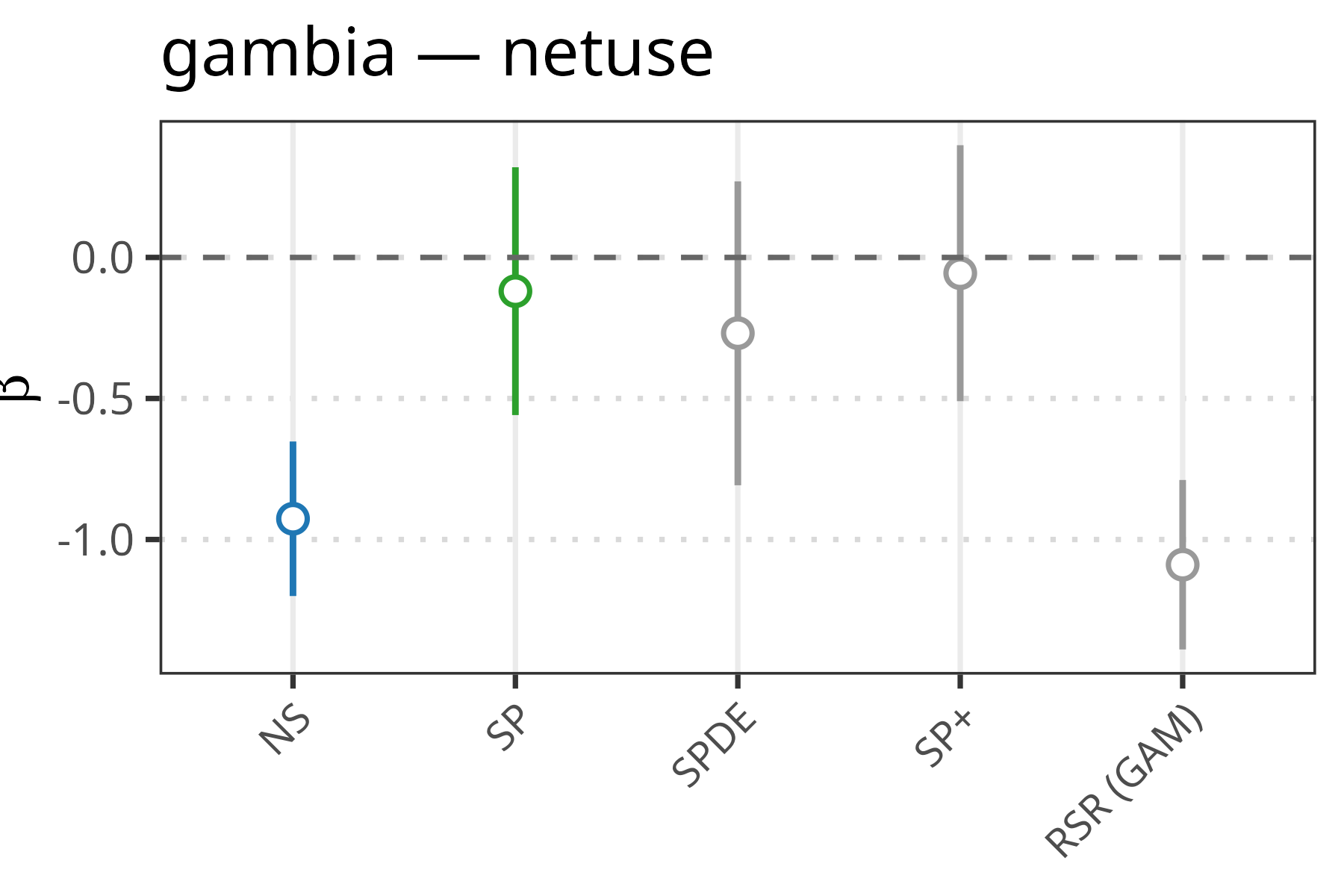}
  \caption{Malaria in Gambia. Estimated effects of (left) vegetation greenness and (right) mosquito net usage across all methods. frequentist methods are shown as point estimates with 95\% confidence intervals and Bayesian methods as posterior means with 95\% credible intervals.}
  \label{fig:gambia_malaria}
\end{figure}

The Gambia malaria dataset~\citep{Thomson1999} consists of point-referenced measurements of malaria incidence, modeled as a function of intervention and environmental covariates such as vegetation greenness and mosquito net usage (Figure~\ref{fig:gambia_malaria}).  

In the non-spatial model, both vegetation greenness and mosquito net usage show clear and statistically significant associations with malaria incidence, with estimates that are consistent with the substantive intuition behind these variables.  
However, once spatial structure is introduced through GLS/PLM, the corresponding coefficients are shrunk towards zero and their 95\% intervals include zero, so that the effects are no longer statistically significant.  
The Spatial$+$ variants follow the same pattern: for both covariates, point estimates remain in the same direction as in the non-spatial model but with reduced magnitude and wider intervals, again leading to non-significant effects.
RSR preservers the significance from the non-spatial model.

\section{Discussion and conclusion}
\label{sec:conclusion}

This paper presents a detailed review of the development of methods to alleviate spatial confounding over the past two decades. Our main goal here is to provide an up-to-date overview of spatial confounding methods, their diversity of applications, and some available software for practitioners.

Although the concept of spatial confounding has been in the literature for a while (as shown in Section~\ref{sec:filter}), the term 'spatial confounding' was coined by \citet{2006_Reich_Biometrics} as a correlation between covariates and the structured random effect in a spatial model, causing bias in the estimates of the linear model.
The first formal proposed solution for the concept was a restricted the structured random effect to the orthogonal complement of the space spanned by the covariates.
This solution has since been shown to be problematic, relying on strong hypotheses and having coverage issues.
Spatial filtering methods were also used to handle spatial confounding in a different literature \citep{2024_Donegan_GA}.

In parallel to the first developments by RSR, the nature of confounding was being unraveled \citep{2010_Paciorek_StatSci}.
The definition of confounding started to shift from a correlation of a random effect, that is, a problem with the model in question, to an omission bias problem. 
The nature of the omitted variable and its relation to the observed variables showed to be crucial to how effective spatial methods are to mitigate bias and when correction for spatial confounding  is necessary.
It has been shown that bias can be effectively mitigated when the observed variables vary at a finer scale than the non-observable ones.
In this context, the Spatial+ method appeared as a promising alternative to mitigate bias \citep{2022_Dupont_Biometrics}. 
Spatial+ filters spatial variation from the observed covariates by adjusting a smooth spatial surface and extracting its residuals, removing indirect effects from any spatially structured covariate confounding the regression.
Although easy to implement, the Spatial+ has a cumbersome step in selecting a basis for fitting the surface.

Discussions on spatial confounding eventually spread to the causality practitioners \citep{2019_Papadogeorgou_Biostatistics, 2021_Reich_ISR,2024_Gilbert_preprint, 2024_Woodward_preprint}.
There, spatial confounding was addressed formally under the causal identification assumptions. 
Those are important contributions because they identify agnostic key assumptions of the model used to analyze if there are conditions to draw causal conclusions about the data.

Much has been developed for spatial confounding, but much must be done. 
First, \citet{2024_Gilbert_preprint} found that spatial confounding methods are extremely sensitive to misspecification, especially violations of linearity and homogeneity. 
Recently, many machine learning methods have been adapted to deal with spatial data \citep[e.g.,][]{sigrist2020Gaussian, zhan2024neural}, but there is still need for careful study of when these methods may fail to correctly estimate treatment effects and how they react to smooth confounders. 
The sensitivity to heterogeneity has also been poorly studied in the literature, but there might be promising connections. 

Spatio-temporal confounding is another topic of interest, but it has not been thoroughly addressed to date. 
Some methods, dating almost two decades, resemble spatial confounding methods that work by scale \citep{janes2007trends}.
The copula structure developed by \citet{2022_Prates_JRSSC} captures spatio-temporal dynamics and has random effects separated from the fixed effects, possibly avoiding confounding bias. 
\citet{2024_Zaccardi} propose Bayesian dynamic spatio-temporal models to deal with what he first named \textit{spatio-temporal confounding}.
In their review, \citet{2021_Reich_ISR} briefly discuss extensions of the methods for spatial-temporal data and Granger causality for this scenario.
\citet{2023_Aritz_StatMod} apply RSR methods to spatio-temporal data. A great advance for spatio-temporal confounding would be an environment with benchmarking datasets and the possibility to synthetically generate confounding for spatio-temporal data, as we have with SpaCE for spatial data \citep{2024_tec_SpaCE}.

Interference is another phenomenon which invalidates many modeling assumptions made in a spatial context.
Also known as a spillover effect, interference happens when the outcome in one location might be driven by exposures in the same and other locations \citep{papadogeorgou2024spatialcausalinferencepresence}.
\citet{1991_Halloran} defined all key estimates in a scenario with interference, in which we highlight the indirect effect of a treatment, that is, a portion of a unit’s effect due to the administration of treatment in other units (nearby ones, for example).
In spatial  econometrics, the indirect effect is of important interest \citep{2009_LeSagePace} and models such as the SAR and Spatial Durbin models are used to recover this indirect effect.
\citet{2021_Reich_ISR} review the challenges associated with the interference assumption.
\citet{papadogeorgou2024spatialcausalinferencepresence} studied spatial confounding in the presence of interference, with the key observation that interference and confounding can manifest as one another.
This is of major interest because, in a situation where we are trying to mitigate confounding, we commonly miss an important covariate, such as the indirect effect.

As the key estimate in spatial confounding is the true effect of a variable on the outcome, quantification of uncertainty becomes overlooked in the literature. 
\citet{2023_Marques_preprint} showed promising directions by proposing a Bayesian version of the Spatial+ method, where uncertainty is naturally addressed.
However, as seen in \citet{2024_Gilbert_preprint}, many popular methods rely on bootstrapping for uncertainty quantification, which may not have good properties under dependent data.
Finally, sensitivity is another topic that has little development.
Many methods depend on a specific choice of basis to express a spatial smooth surface.
However, no established method exists to study such sensitivity, and this is a promising area for further research.






\begin{acks}[Acknowledgments]
The Coordination for the Improvement of Higher Education Personnel (CAPES) and Fundação Getulio Vargas (FGV) for financial support. Marcos O. Prates acknowledges the research grants obtained from CNPq-Brazil (309186/2021-8), FAPEMIG (APQ-01837-22, APQ-01748-24), and CAPES, respectively, for partial financial support.
\end{acks}

\bibliographystyle{imsart-nameyear} 
\bibliography{sp_confounding}       

\end{document}